\PassOptionsToPackage{table}{xcolor}

\documentclass[11pt]{article}
\usepackage[margin=1in]{geometry}
\usepackage{cprotect}
\usepackage{color}
\PassOptionsToPackage{hyphens}{url}\usepackage[colorlinks=false,linkcolor=blue,citecolor=blue,urlcolor=blue]{hyperref}
\usepackage{amssymb,amsmath}
\usepackage{amsthm}
\usepackage{graphicx}
\usepackage{helvet}
\usepackage[font=footnotesize,labelfont=bf]{caption}
\usepackage{subcaption}
\usepackage{longtable}

\usepackage{multirow}
\usepackage{tabularx}
\usepackage{booktabs}
\usepackage{diagbox}
\usepackage[demo]{rotating}
\usepackage{titlesec}
\usepackage[utf8]{inputenc}
\usepackage{pifont}
\usepackage{array, makecell} %
\usepackage{bm}
\titleformat*{\section}{\LARGE\bfseries}
\titleformat*{\subsection}{\Large\bfseries}
\titleformat*{\subsubsection}{\large\bfseries}
\usepackage{rotating}
\usepackage{tikz}
\usepackage[graphicx]{realboxes}

\usepackage{xr}
\usepackage[square,sort&compress,numbers]{natbib}
\makeatletter
\def\blfootnote{\xdef\@thefnmark{}\@footnotetext}
\makeatother

\usepackage{setspace}
\usepackage{pifont}
\newcommand{\checkcross}{\checkmark\kern-1.1ex\raisebox{.7ex}{\rotatebox[origin=c]{125}{--}}}

\numberwithin{equation}{section}
\theoremstyle{plain}

\usepackage{mathtools}

\usepackage{tabulary}
\usepackage{multirow}
\usepackage{float}
\usepackage{array}
\newcolumntype{L}[1]{>{\raggedright\let\newline\\\arraybackslash\hspace{0pt}}m{#1}}
\newcolumntype{C}[1]{>{\centering\let\newline\\\arraybackslash\hspace{0pt}}m{#1}}
\newcolumntype{R}[1]{>{\raggedleft\let\newline\\\arraybackslash\hspace{0pt}}m{#1}}

\usepackage{xcolor}

\usepackage{amssymb}
\usepackage{pifont}

\newcommand{\hcmark}{\textcolor{black}{\ding{51}}{\small\textcolor{black}{\kern-0.7em\ding{55}}}}
\definecolor{myblue}{RGB}{30, 112, 161}
\newcommand{\rv}[1]{\textcolor{black}{#1}}
\newcommand{\p}{{\rm I}\kern-0.18em{\rm P}}
\newcommand{\1}{{\rm 1}\kern-0.24em{\rm I}}

\newcommand{\E}{\mathbb{E}}

\newcommand{\Var}{\mathrm{Var}}

\usepackage{tikz}

\title{Categorization of \rv{33} computational methods to detect spatially variable genes from spatially resolved transcriptomics data}
\author{Guanao Yan\,$^{1}$, Shuo Harper Hua\,$^{2}$, and Jingyi Jessica Li\,$^{1,3,4,5,6,*}$
}

\date{\vspace{-5ex}}

\begin{document}

\maketitle

\blfootnote{\\$^{1}$ Department of Statistics, University of California, Los Angeles, CA 90095-1554\\
$^{2}$ Department of Biomedical Data Science, Stanford University, Stanford, CA 94305\\
$^{3}$ Department of Human Genetics, University of California, Los Angeles, CA 90095-7088\\
$^{4}$ Department of Computational Medicine, University of California, Los Angeles, CA 90095-1766\\
$^{5}$ Department of Biostatistics, University of California, Los Angeles, CA 90095-1772\\
$^{6}$ Radcliffe Institute for Advanced Study, Harvard University, Cambridge, MA 02138\\
$^{*}$ To whom correspondence should be addressed. Email: jli@stat.ucla.edu}

\begin{abstract}

In the analysis of spatially resolved transcriptomics data, detecting spatially variable genes (SVGs) is crucial. Numerous computational methods exist, but varying SVG definitions and methodologies lead to incomparable results. We review \rv{33} state-of-the-art methods, categorizing SVGs into three types: overall, cell-type-specific, and spatial-domain-marker SVGs. Our review explains the intuitions underlying these methods, summarizes their applications, and categorizes the hypothesis tests they use in the trade-off between generality and specificity for SVG detection. We discuss challenges in SVG detection and propose future directions for improvement. Our review offers insights for method developers and users, advocating for category-specific benchmarking.

\end{abstract}

\newpage
\section{Introduction}{\label{sec:intro}}
In multicellular organisms, cells work cooperatively in tissues and organs to fulfill biological functions. Measuring cells' spatial locations along with transcriptome profiles can help unravel collaborative cell organizations and molecular mechanisms in tissues and organs \cite{Longo2021}. Spatially resolved transcriptomics (SRT) technologies have been rapidly evolving to enable high-throughput profiling of transcriptomes at spatial locations, which may represent single cells or groups of cells, in a tissue slice. SRT data provide unprecedented insights into genes' spatial expression patterns, tissue's cellular organizations, and cell-cell communications \cite{marx2021method}.

\rv{To date, major SRT technologies are either imaging-based or sequencing-based, and the two types of technologies have complementary advantages \cite{Longo2021}. 
Imaging-based SRT technologies use fluorescence in situ hybridization (FISH) to measure the expression levels of selected 200--400 target genes at the single-cell or subcellular spatial resolution \cite{Longo2021}. Examples of imaging-based SRT technologies include in situ sequencing (ISS) \cite{Ke2013}, sequential fluorescence in situ hybridization (seqFISH) \cite{lubeck2014single, shah2016situ, eng2019transcriptome}, multiplexed error-robust fluorescence in situ hybridization (MERFISH) \cite{chen2015spatially}, STARmap \cite{wang2018three}, ExSeq \cite{alon2021expansion}, and 10x Xenium \cite{janesick2023high}. The high spatial resolution of these approaches makes them particularly useful for studies requiring precise localization of transcripts within tissue sections. For example, MERFISH is widely used to annotate cell types in subregions of the mouse brain \cite{zhang2021spatially,moffitt2018molecular} and to investigate the spatial interaction between specific cell types in the mouse somatosensory cortex \cite{stogsdill2022pyramidal}. However, imaging-based SRT technologies cannot provide transcriptome-wide gene expression levels.}

\rv{In contrast, sequencing-based SRT technologies, such as Spatial Transcriptomics \cite{shah2016situ}, 10x Visium \cite{Visium}, Slide-seq \cite{rodriques2019slide}, and GeoMx \cite{zollinger2020geomx}, capture transcriptome-wide gene expression at a lower spatial resolution. Each spot's diameter ranges between $10~\mathrm{\mu m}$ and $100~\mathrm{\mu m}$, containing multiple cells of possibly different types \cite{Longo2021}. Recently developed sequencing-based SRT technologies \cite{chen2022spatiotemporal, cho2021microscopic} can achieve higher, even single-cell, spatial resolution, with spot diameters reduced to below $1~\mathrm{\mu m}$, despite the significantly higher cost. Sequencing-based SRT technologies
enable transcriptome-wide analyses, such as identifying novel marker genes for tissue layers \cite{maynard2021transcriptome}, and sequencing-read analysis, such as inferring RNA velocity from spliced and unspliced sequencing reads \cite{abdelaal2021sirv}.}

Although imaging-based and sequencing-based SRT technologies have different spatial resolutions, for simplicity, we use the same term ``spots'' to refer to the measured spatial locations for both technologies in this review.
Hence, SRT data encompass two components: (1) an expression count matrix of $p$ genes at $n$ spots and (2) a spatial coordinate matrix containing the two-dimensional (2D) coordinates of $n$ spots. 

\begin{figure}[htbp]
    \centering
    \includegraphics[width=\textwidth]{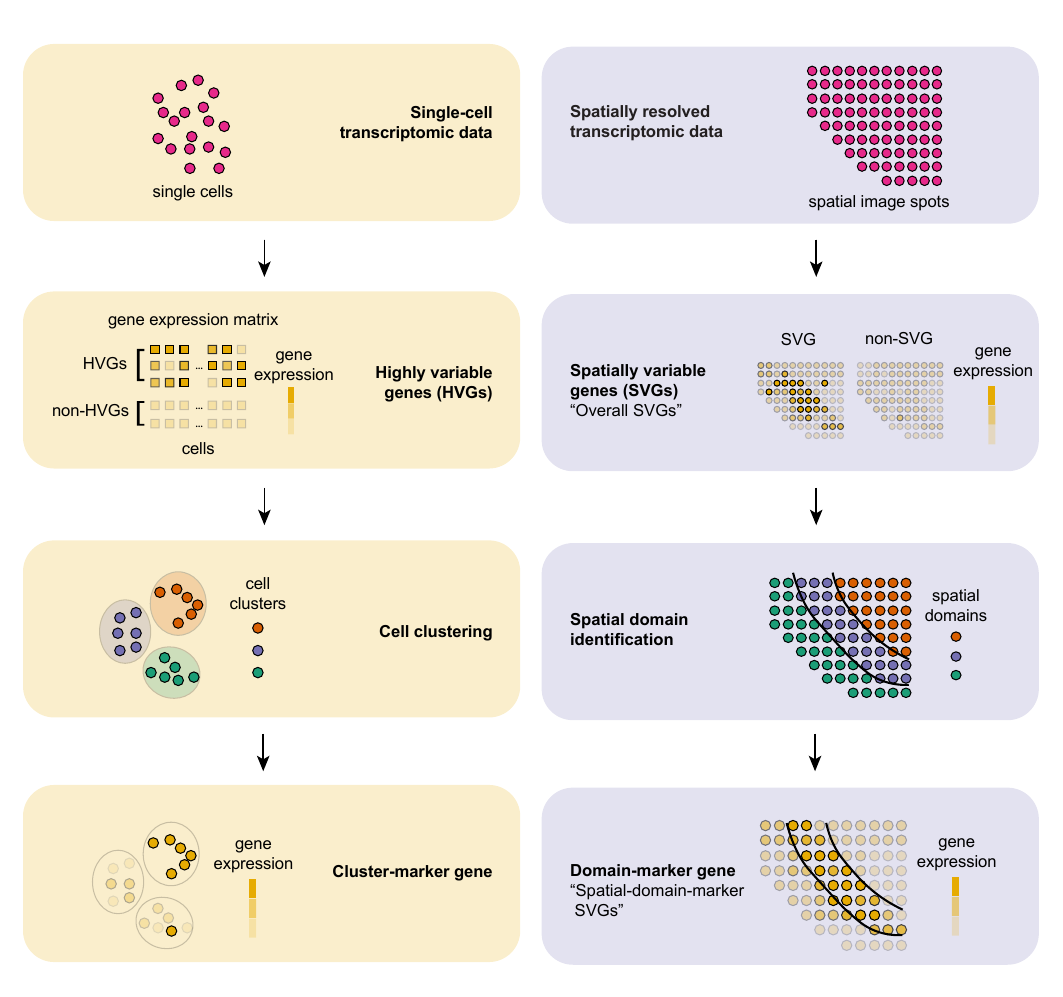}
    \caption{
    \textbf{General analysis workflows of single-cell transcriptomic data and spatially resolved transcriptomic (SRT) data.} The left column shows a general analysis workflow for single-cell transcriptomic data with steps including highly variable gene (HVG) detection, cell clustering, and cluster-marker gene identification. The right column illustrates \rv{a} workflow for analyzing SRT data with steps including spatially variable gene (SVG) detection, spatial domain identification, and domain-marker gene identification.
    }
    \label{fig:pipeline}
\end{figure}

As with other types of high-throughput data where hundreds or thousands of genes are measured simultaneously, a crucial early step of SRT data analysis is the identification of informative genes. Prior to SRT data, single-cell RNA-seq data contain gene expression profiles of single cells without spatial information.
In typical single-cell RNA-seq data analysis (Fig.~\ref{fig:pipeline}), highly variable gene (HVG) detection methods are used to screen a proportion of genes (e.g., $10\%$--$20\%$) with the largest variances (adjusting for cell library sizes in several methods) to reduce the dataset's dimensionality (from $p$ genes to a smaller number of HVGs) and remove the unimportant variations of many genes. The underlying assumption of HVG detection is that genes exhibiting significant expression variations across single cells are more likely to reflect biological variations rather than technical variations caused by sampling effects in sequencing. Analogous to single-cell RNA-seq data analysis, SRT data analysis typically includes an early step to detect \textit{spatially variable genes (SVGs)}, which conceptually generalize HVGs by including the spatial information (Fig.~\ref{fig:pipeline}). Intuitively, SVGs are genes whose gene expression levels exhibit non-random, informative spatial patterns.  

In single-cell RNA-seq data analysis (Fig.~\ref{fig:pipeline}), common steps after HVG detection include cell clustering and differentially expressed gene (DEG) detection. The goal of cell clustering is to identify potential cell types, and subsequent DEG detection aims to find the genes that are \rv{significantly more highly} expressed in each cluster. The resulting \rv{highly-expressed DEGs of each cell cluster} are then used to annotate the cluster as a particular cell type (or subtype), serving as cell-type markers. Similarly, in SRT data analysis, following SVG detection, spots are often partitioned into spatial domains, where each domain contains proximal spots exhibiting similar gene expression profiles. Parallel to DEG detection between clusters in single-cell RNA-seq data analysis, DEGs can be identified between spatial domains, and the resulting \rv{highly-expressed DEGs of each spatial domain} serve as spatial-domain markers.

Many computational methods have been developed to detect SVGs. However, the SVG definitions in these methods lack consensus, resulting in diverse meanings for the detected SVGs and making understanding difficult. Consequently, the inconsistent definitions of SVGs may lead to ambiguous usage of computational methods. Although 19 methods were reviewed previously \cite{adhikari2024recent}, the review did not categorize SVG definitions but only focused on summarizing the methodologies, including the input data type (count data vs. normalized data), algorithm type (model-based vs. model-free), statistical paradigm (frequentist vs. Bayesian), the availability of false discovery rate (FDR) control, etc. Nor did the review discuss the biological implications and downstream applications of SVGs.

\begin{figure}[htbp]
    \centering
    \includegraphics[width=\textwidth]{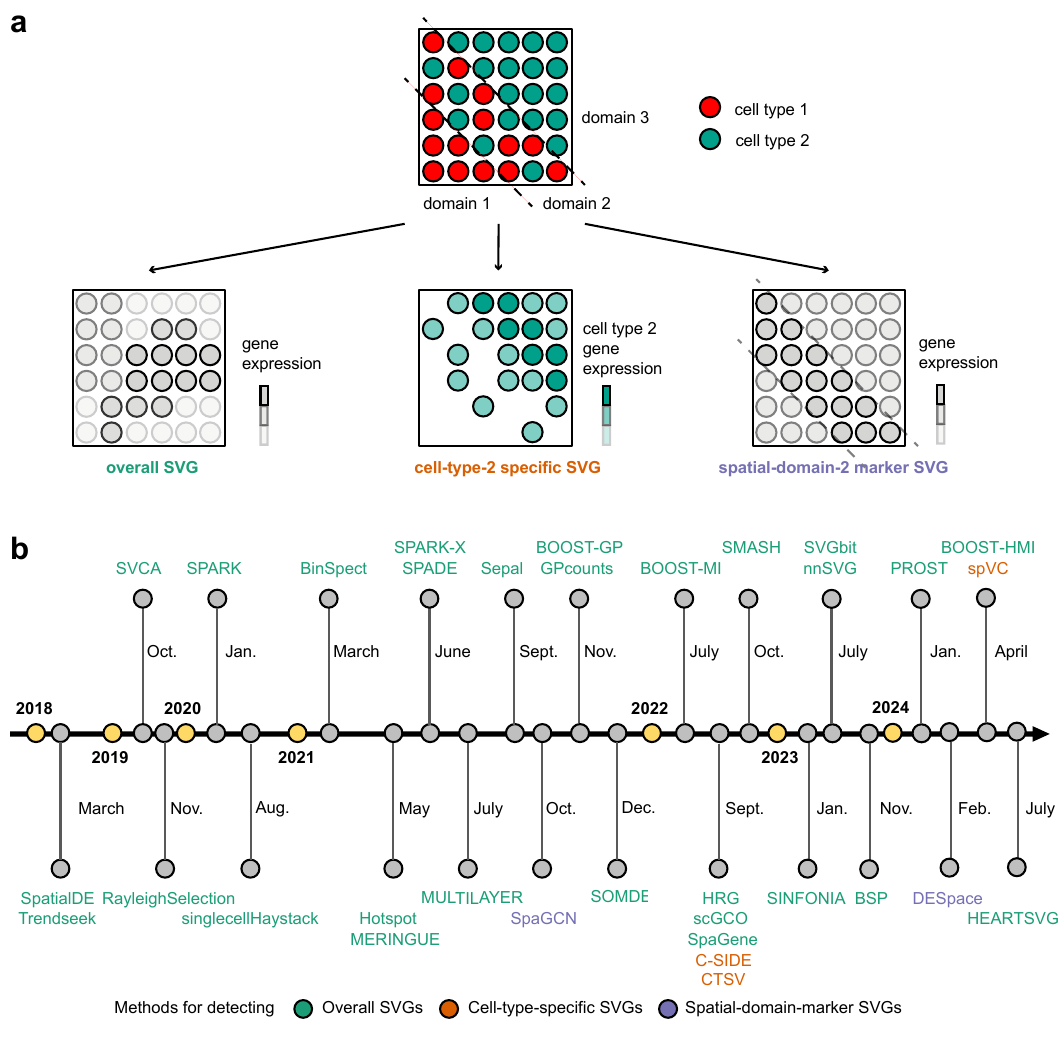}
    \caption{
    \textbf{Conceptual visualization of three SVG categories and timeline of 33 SVG detection methods.} \textbf{a.} Conceptual visualization of three SVG categories: overall SVGs, cell-type-specific SVGs, and spatial-domain-marker SVGs. The top row shows a tissue slice with two cell types and three spatial domains. From left to right, exemplar genes with colors representing the expression levels are shown for an overall SVG, a cell-type-specific SVG, and a spatial-domain-marker SVG, respectively. 
    \textbf{b.} Publication timeline of 33 SVG detection methods. Colors represent three SVG categories: overall SVGs (green), cell-type-specific SVGs (red), and spatial-domain-marker SVGs (purple).
    }
    \label{fig:catecartoon}
\end{figure}

\begin{figure}[htbp]
    \centering
    \includegraphics[width=\textwidth]{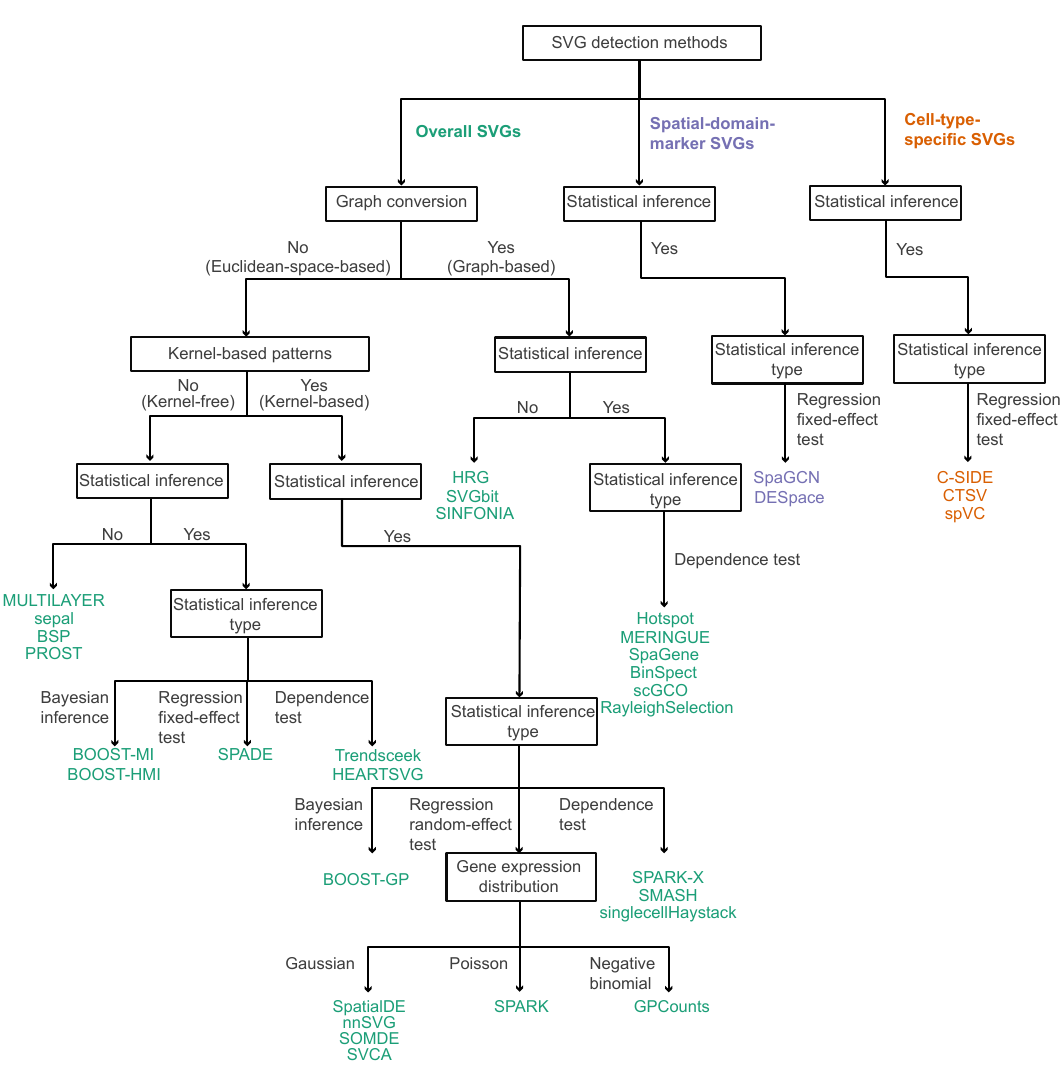}
    \caption{
    \textbf{A hierarchical summary of 33 SVG detection methods.} The hierarchical summary considers the methodological characteristics, including graph conversion, kernel-based patterns, availability of statistical inference, statistical inference types, and gene expression distributions. Colors represent three SVG categories: overall SVGs (green), cell-type-specific SVGs (red), and spatial-domain-marker SVGs (purple).
    }
    \label{fig:decisiontree}
\end{figure}

Motivated by this gap in understanding SVGs, here we review \rv{33} peer-reviewed SVG detection methods (Fig.~\ref{fig:catecartoon}) and define three SVG categories: \textit{overall SVGs}, \textit{cell-type-specific SVGs}, and \textit{spatial-domain-marker SVGs}. 
For the first time, our review unveils the biological significance underpinning the three categories of SVGs, summarizes the frequentist hypothesis tests implemented in \rv{23} SVG detection methods, and
discusses the limitations of existing methods and outlines future directions for improvement. 
Moreover, we construct a hierarchy to summarize the \rv{33} SVG detection methods in terms of methodological differences in Fig.~\ref{fig:decisiontree} and list the technical details in Tables~\ref{tab:cate}--\ref{tab:tab_test}. \rv{Additionally, to help readers better understand our review, we provide Table~\ref{tab:concepts_summary}, which lists the definitions of the frequently used terminologies.}

\rv{Methods for detecting the three SVG categories serve different purposes (Fig.~\ref{fig:catecartoon}a). First, the detection of overall SVGs screens informative genes for downstream analysis, including the identification of spatial domains and functional gene modules. Second, detecting cell-type-specific SVGs aims to reveal spatial variation within a cell type and help identify distinct cell subpopulations or states within cell types. Third, spatial-domain-marker SVG detection is used to find marker genes to annotate and interpret spatial domains already detected. These markers help understand the molecular mechanisms underlying spatial domains and assist in annotating tissue layers in other datasets.} 

The relationship among the three SVG categories depends on the detection methods, particularly the null and alternative hypotheses they employ. If an overall SVG detection method uses the null hypothesis that a non-SVG's expression is independent of spatial location and the alternative hypothesis that any deviation from this independence indicates an SVG, then its SVGs should theoretically include both cell-type-specific SVGs and spatial-domain-marker SVGs. \rv{For example, DESpace \cite{cai2024despace} is a method that detects both overall SVGs and spatial-domain-marker SVGs, and its detected overall SVGs must be marker genes for some spatial domains.} This inclusion relationship holds true except in extreme scenarios, such as when a gene exhibits opposite cell-type-specific spatial patterns that effectively cancel each other out. However, if an overall SVG detection method's alternative hypothesis is defined for a specific spatial expression pattern, then its SVGs may not include some cell-type-specific SVGs or spatial-domain-marker SVGs. \rv{For readers with a statistical background seeking deeper insights,} we provide a comprehensive discussion on the frequentist hypothesis tests implemented by \rv{23} SVG detection methods in the Section ``Theoretical characterization of SVG detection methods that use frequentist hypothesis tests." \rv{Other readers can skip this section and still grasp the core content of our review.}

Although three benchmark studies were conducted to compare SVG detection methods \cite{charitakis2023disparities, chen2024evaluating, li2023benchmarking}, they have three major differences from our review. First, the benchmark studies performed numerical comparisons of SVG detection methods, while our review focuses on categorizing methods conceptually and methodologically. Second, the benchmark studies did not categorize SVGs but focused on detecting the overall SVGs by our definition. For example, spaGCN, a method that detects spatial-domain-marker SVGs, does not perform well in a benchmark study \cite{li2023benchmarking}, possibly because the study focused on detecting overall SVGs. This observation underscores the need to categorize SVGs. Third, our review has a more comprehensive coverage of SVG detection methods: \rv{33} peer-reviewed methods (all methods to our knowledge) compared to \rv{6 methods in Charitakis \textit{et al.}~\cite{charitakis2023disparities}, 7 methods in Chen \textit{et al.}~\cite{chen2024evaluating}, and 14 methods (12 peer-reviewed) in Li \textit{et al.}~\cite{li2023benchmarking}}. In summary, existing benchmark studies complement our review by providing numerical evidence to rank methods in detecting certain types of overall SVGs. Future benchmark studies are needed to compare methods within the three SVG categories and examine more types of null and alternative hypotheses.

\rv{To facilitate our discussion, we introduce the following mathematical notations. For each gene with expression levels (in counts or normalized) measured at \( n \) spatial spots, we use \( y_{i} \in \mathbb{R} \) represent its expression level at spot \( i = 1, \ldots, n \). When gene expression levels are considered random variables, we use \( Y_{i} \) to emphasize the random nature. The vector notation is $\mathbf{Y}=(Y_1,\ldots,Y_n)^\top$. For each spot \(i = 1, \ldots, n\), we denote its 2D spatial location as \(\mathbf{s}_i = (s_{i1}, s_{i2})^\top \in \mathbb{R}^{2}\). The spatial coordinates of all \(n\) spots are represented by the matrix \(\mathbf{s} = \left[\mathbf{s}_{1}, \ldots, \mathbf{s}_{n}\right]^{\top} \in \mathbb{R}^{n \times 2}\), where each row corresponds to a spot. When spots are annotated with \(L\) spatial domain labels, the spatial-domain indicator vector for spot \(i\) is \(\mathbf{d}_i = (d_{i1}, \ldots, d_{iL})^{\top} \in \{0, 1\}^{L}\), with \(\sum_{l=1}^{L} d_{il} = 1\); that is, \(d_{il} = 1\) means that spot \(i\) belongs to spatial domain \(l\). When spots are annotated with \(K\) cell type labels, the cell-type proportion vector for spot \(i\) is \(\mathbf{c}_i = (c_{i1}, \ldots, c_{iK})^{\top} \in [0, 1]^{K}\), where \(c_{ik}\) indicates the proportion of cell type \(k\) at spot \(i\), with \(\sum_{k=1}^{K} c_{ik} = 1\). In summary, each spot \(i\) has three covariate vectors: spatial location \(\mathbf{s}_i\), spatial-domain indicator \(\mathbf{d}_i\), and cell-type proportions \(\mathbf{c}_i\). We summarize the mathematical notations in Table~\ref{tab:notations}.}

\section{Methods for detecting overall SVGs}\label{sec:SVGMeth_HVG}

\rv{We define \textit{overall SVGs} as the genes that exhibit non-random spatial expression patterns. This is the most general category of SVGs detected using only SRT data without incorporating external information such as spatial domains or cell types (See Fig.~\ref{fig:catecartoon}a and Table~\ref{tab:concepts_summary}).} Methods for detecting overall SVGs are generally classified into Euclidean-space-based and graph-based methods (Fig.~\ref{fig:decisiontree}). Euclidean-space-based methods analyze spots in a tissue slice by considering their locations in a 2D Euclidean space. In contrast, graph-based methods first construct a graph of spatial spots and then analyze this graph, focusing on the connections between spots rather than their Euclidean distances. Both approaches, regardless of the spatial representation they use (Euclidean space or graph), study how a gene's expression levels vary across spatial spots, examining the relationship between each gene and its spatial context.


\subsection{Euclidean-space-based methods}
Euclidean-space-based methods are further divided based on whether they use a kernel function to target a specific spatial pattern. Kernel-based methods utilize a kernel function to specify the covariances of spatial spots, enhancing their power to detect the targeted spatial patterns. In contrast, kernel-free methods do not use a kernel function and instead rely on other approaches to capture spatial patterns.

\subsubsection{Kernel-based methods}

Kernel-based methods use a pre-defined kernel function \( K(\mathbf{s}_i, \mathbf{s}_j) \) to specify the covariance between spots \( i \) and \( j \) based on their spatial locations. This kernel function usually decays as the Euclidean distance between spots \( i \) and \( j \) increases. Define \(\mathbf{K}(\mathbf{s}) = [K(\mathbf{s}_i, \mathbf{s}_j)]\) as the \(n \times n\) matrix representing these covariances. The most commonly used kernel functions include the Gaussian kernel for detecting clustered or focal expression patterns, expressed as
\begin{equation}\label{eq:Gaussian_kernel}
    K_G(\mathbf{s}_i, \mathbf{s}_j)=\exp \left(-\frac{\left\|\mathbf{s}_i-\mathbf{s}_j\right\|^2}{2 \sigma^2}\right)\,,
\end{equation}
and the cosine kernel for detecting periodic expression patterns, denoted by
\begin{equation}\label{eq:cosine_kernel}
    K_C(\mathbf{s}_i, \mathbf{s}_j)=\cos \left(2 \pi \frac{\left\|\mathbf{s}_i-\mathbf{s}_j\right\|}{\phi}\right)\,.
\end{equation}
For a given gene, the covariance matrix of its expression vector $\mathbf{Y}$ (may be subject to transformation) can be decomposed into components, one of which depends on \(\mathbf{K}(\mathbf{s})\). If the contribution of this term is significantly positive according to a hypothesis test, commonly known as the ``variance component test," the gene is detected as an overall SVG.

Following this general idea, \rv{ten} kernel-based methods (Fig.~\ref{fig:decisiontree}) make different assumptions about the distribution of \(\mathbf{Y}\) and the covariance matrix decomposition. Below, we briefly summarize these methods and highlight their similarities and differences.

\textbf{SpatialDE} \cite{svensson2018spatialde} assumes that a gene's normalized expression \(\mathbf{Y} \in \mathbb{R}^n \) follows an \(n\)-dimensional Gaussian distribution, with a covariance matrix that includes a spatial covariance component involving \(\mathbf{K}(\mathbf{s})\):
\begin{eqnarray*}
    \mathbf{Y} \sim \operatorname{MVN}\left(\bm{\mu},\; {\sigma}_s^2 \cdot \mathbf{K}(\mathbf{s})+\delta \cdot \mathbf{I}\right),
\end{eqnarray*}
where \(\operatorname{MVN}\) represents a multivariate Gaussian distribution with a mean vector \(\bm{\mu} \in \mathbb{R}^n\) and a covariance matrix \({\sigma}_s^2 \cdot \mathbf{K}(\mathbf{s}) + \delta \cdot \mathbf{I}\). Here, \(\mathbf{I}\) is an \(n\)-dimensional identity matrix, and \(\sigma_s^2\) and \(\delta\) are parameters representing the spatial covariance component and the error variance component, respectively. This model is a realization of a Gaussian process with the kernel function $K(\cdot,\cdot)$. To determine if a gene is an overall SVG, SpatialDE employs a likelihood ratio test with the null hypothesis $H_0: \sigma_s^2 = 0$, comparing this model with a null model that does not include the spatial covariance component: $\mathbf{Y} \sim \operatorname{MVN}\left(\bm{\mu},\; {\sigma}^2 \cdot \mathbf{I}\right)$.

Using the same model as SpatialDE, \textbf{nnSVG} \cite{weber2023nnsvg} improves computational efficiency by replacing the Gaussian process with the nearest-neighbor Gaussian process \cite{datta2016hierarchical}, providing a scalable approximation. This enhancement allows nnSVG's computational complexity and runtime to scale linearly with \( n \), the number of spatial spots, rather than cubicly.

\textbf{SOMDE} \cite{hao2021somde} also follows the statistical model of SpatialDE but adds a data preprocessing step for scalability. Specifically, SOMDE condenses the original spatial spots into fewer grid points and assigns each grid point a ``meta-expression," which aggregates a gene's expression levels at the spots condensed into that grid point. After this preprocessing step, SOMDE uses the same approach as SpatialDE to detect whether the gene is an overall SVG, but it does so using the grid points instead of the original spots.

Compared to the first three methods, \textbf{SVCA} \cite{arnol2019modeling} modifies the covariance matrix decomposition by adding two additional components:
\[
    \mathbf{Y} \sim \operatorname{MVN}\left(\bm{\mu},\; \mathbf{K}_{\text{int}} + \mathbf{K}_{\text{c-c}} + {\sigma}_s^2 \cdot \mathbf{K}(\mathbf{s}) + \sigma^2 \cdot \mathbf{I}\right),
\]
where \(\mathbf{K}_{\text{int}}\) denotes the intrinsic cell-state covariance, and \(\mathbf{K}_{\text{c-c}}\) represents the cell-cell interaction covariance. Specifically, \(\mathbf{K}_{\text{int}}\) is defined solely using gene expression data without spatial information, while \(\mathbf{K}_{\text{c-c}}\) incorporates both gene expression data and Euclidean distances among spots.

Instead of using normalized gene expression data, \textbf{SPARK} \cite{sun2020statistical}, \textbf{GPcounts} \cite{bintayyashNonparametricModellingTemporal2021}, and \textbf{BOOST-GP} \cite{liBayesianModelingSpatial2021} directly model a gene's expression count at spot \(i\), \(Y_i \in \mathbb{N}\), using Poisson, negative binomial (NB), and zero-inflated negative binomial (ZINB) distributions, respectively. For instance, \textbf{SPARK} assumes that 
\[
    Y_i \overset{\mathrm{ind}}{\sim} \operatorname{Poisson}\left( \mu_i(\mathbf{s}_i)\right),~ i=1,2, \cdots, n\,.
\]
For spot \(i\), the Poisson mean parameter \(\mu_i(\mathbf{s}_i)\) is a function of the spatial location \(\mathbf{s}_i\), specified by
\[
    \log \left(\mu_i(\mathbf{s}_i)\right)=\mathbf{x}_i^\top \bm{\beta}+b_i(\mathbf{s}_i)+\epsilon_i\,,
\]
where \(\mathbf{x}_i\) indicates spot \(i\)'s covariates, and \(b_i(\mathbf{s}_i)\) is the random intercept at spot \(i\). The \(n\) random intercepts, \(b_1(\mathbf{s}_1), \ldots, b_n(\mathbf{s}_n)\), are assumed to follow a Gaussian process with the kernel function \(K(\cdot, \cdot)\):
\[
    \left(b_1(\mathbf{s}_1), b_2(\mathbf{s}_2), \cdots, b_n(\mathbf{s}_n)\right)^\top \sim \operatorname{MVN}\left(0,\; \sigma_s^2 \cdot \mathbf{K}(\mathbf{s})\right).
\]
To decide whether the gene is an overall SVG, SPARK tests the null hypothesis \(H_0: \sigma_s^2 = 0\). Thus, SPARK shares the same variance component test idea with SpatialDE but uses a hierarchical model, incorporating a Gaussian process in the top layer and Poisson distributions in the bottom layer, to account for count data.

Unlike the previous \rv{seven} methods that use parametric models, \rv{three} methods---\rv{\textbf{SPARK-X} \cite{zhu2021spark}, \textbf{SMASH} \cite{seal2023smash}, and \textbf{singlecellHaystack} \cite{vandenbon2020singlecellHaystack}}---employ non-parametric tests to determine whether a gene's expression is independent of its spatial location. 

\textbf{SPARK-X} \cite{zhu2021spark} tests whether two \(n \times n\) spot similarity matrices are independent to decide if a gene is an overall SVG. 
\rv{One similarity matrix is based on the gene's expression levels at the \(n\) spots, defined as $\mathbf{E}=\mathbf{Y}\left(\mathbf{Y}^\top \mathbf{Y}\right)^{-1} \mathbf{Y}^\top$. The other is based on the kernel-transformed spatial locations of the \(n\) spots, defined as $\mathbf{K}=\mathbf{s'}\left((\mathbf{s}')^\top \mathbf{s}'\right)^{-1} (\mathbf{s}')^\top$, where $\mathbf{s}'$ is an $n \times 2$ matrix whose $i$-th row $\mathbf{s}_i'=(s_{i1}', s_{i2}')^\top$ encode the kernel-transformed spatial location of spot $i=1,\ldots,n$.}
Specifically, SPARK-X transforms the spatial location \(\mathbf{s}_i = (s_{i1}, s_{i2})^\top\)  of spot $i$ using two kernel-based functions: a Gaussian transformation \(s_{il}' = \exp\left(\frac{-s_{il}^2}{2 \sigma_l^2}\right)\), \(l = 1, 2\), to detect clustered or focal patterns, and a cosine transformation \(s_{il}' = \cos\left(\frac{2 \pi s_{il}}{\phi_l}\right)\), \(l = 1, 2\), to detect periodic patterns, where \(\sigma_1\), \(\sigma_2\), \(\phi_1\), and \(\phi_2\) are tuning parameters. \rv{Eventually, SPARK-X uses the Pearson correlation between the two similarity matrices, $\operatorname{trace}(\mathbf{E}\mathbf{K})/n$, as the test statistic to decide if a gene is an overall SVG.}

\rv{\textbf{SMASH} \cite{seal2023smash} generalizes SPARK-X by adding two alternatives to the spatial location similarity matrix $\mathbf{K}$. The two alternatives include the Gaussian kernel covariance matrix $\mathbf{K}_G=[K_G(\mathbf{s}_i, \mathbf{s}_j)]$ and the cosine kernel covariance matrix $\mathbf{K}_C=[K_C(\mathbf{s}_i, \mathbf{s}_j)]$, where $K_G(\cdot,\cdot)$ and $K_C(\cdot,\cdot)$ are the kernel functions defined in \eqref{eq:Gaussian_kernel} and \eqref{eq:cosine_kernel}, respectively. Then, SMASH uses the same test statistic as SPARK-X.}

\textbf{singlecellHaystack} \cite{vandenbon2020singlecellHaystack} is a test of independence between a gene's expression level and its spatial location. It involves two pre-processing steps: first, the gene's expression levels at various spots are binarized into two states: detected and undetected; second, the 2D Euclidean space of a tissue slice is divided into a grid along both axes, with intersection points defined as grid points, serving as coarse spatial coordinates. After pre-processing, the method tests if the gene's two expression states are randomly distributed across the grid points. If they are not, the gene is considered an overall SVG. singlecellHaystack uses a Gaussian kernel to define three distributions of grid points: a reference distribution based on all spatial spots, a conditional distribution based on spatial spots in the detected state, and another conditional distribution based on spatial spots in the undetected state. Technically, for a grid point with a 2D spatial location \(\mathbf{s}=(s_1, s_2)^\top\), its reference distribution density is defined as \(\sum_{i=1}^n \exp\left( -\frac{\left\|\mathbf{s}_i-\mathbf{s}\right\|^2}{2} \right)\), subject to normalization. The two conditional distribution densities are similarly defined, with the sum across all spots replaced by sums across the detected and undetected spots, respectively. The test statistic is defined as the sum of two Kullback-Leibler (KL) divergences, each representing the deviance of one conditional distribution from the reference distribution. Intuitively, the larger the test statistic, the more likely the gene is an overall SVG. Finally, the test statistic is converted to a p-value using a permutation test, which shuffles the gene's expression levels across spots.

To summarize, all \rv{ten} kernel-based methods require pre-specification of the kernel function \(K(\cdot, \cdot)\) to target specific spatial patterns. Consequently, the statistical power of these methods depends on how well the kernel function captures the spatial expression patterns of biologically informative genes in the SRT data. 
Due to the limited choices and subjectivity of kernel functions, these methods may be insufficient for detecting complex spatial expression patterns, such as those found in cancer SRT data.

\subsubsection{Kernel-free methods}

\rv{Nine} kernel-free methods employ diverse approaches to detect overall SVGs without relying on pre-specified kernel functions (Fig.~\ref{fig:decisiontree}). Instead, they utilize various statistical and computational techniques to capture spatial patterns. Below, we briefly summarize each method.

\textbf{Trendsceek} \cite{edsgard2018trendsceek} tests if a gene's expression level is dependent of the spatial location using a marked point process, which models the joint probability distribution of spatial locations $\mathbf{s}_i$, $i=1,\ldots,n$, as ``points" and a given gene's expression levels $y_i$, $i=1,\ldots,n$, as ``marks." If deemed dependent, the gene is detected as an overall SVG. For a pair of points $(\mathbf{s}_i, \mathbf{s}_j)$ and their corresponding marks $(y_i, y_j)$, Trendsceek parameterizes their joint probability density by the distance (called "radius") $r_{ij} = |\mathbf{s}_i - \mathbf{s}_j|$ and the marks $y_i$, $y_j$ as $f(y_i, y_j, r_{ij})$. Formally, the dependence of the gene's expression on spatial location is formulated as a conditional density function given the radius $r_{ij} = r$:
\[
    g(r) = M(y_i, y_j | r_{ij} = r) = \frac{f(y_1, y_2, r)}{f(r)}\,.
\]
The univariate function $g(\cdot)$ can then be summarized into four mark-segregation summary statistics as functions of the radius $r$, such as Stoyan's mark-correlation function. To test if the gene is an overall SVG, Trendsceek implements a permutation test by sampling marks without replacement and randomly reassigning them to points. For each radius $r$, summary statistics are calculated and compared to the null distributions derived from the permutations to obtain p-values, which are then combined into a single p-value across all radius values.  \rv{Note that Trendsceek has inspired the ``markvariogram" approach used in the single-cell and SRT analysis toolkit Seurat \cite{seurat} for SVG detection.}

\textbf{MULTILAYER} \cite{moehlin2021multilayer}, similar to singlecellHaystack, discretizes spatial spots for each gene based on the gene's expression levels. The discretization procedure in MULTILAYER consists of three steps. First, each spot receives a $\log_2$ fold change, defined as the ratio of the gene's normalized expression level at the spot to its average normalized expression level across all spots. The $\log_2$ fold changes with absolute values greater than $1$ are truncated to $1$ with the corresponding sign. Second, MULTILAYER applies agglomerative clustering to the spots based on the $\log_2$ fold change values, using a pre-specified distance threshold. Third, if a cluster contains contiguous spots with all positive $\log_2$ fold change values, MULTILAYER assigns the cluster a label of ``1." Finally, MULTILAYER uses the size of the largest cluster labeled as ``1" as a summary statistic for the gene and ranks genes from high to low based on this statistic. Thus, if a gene has a large contiguous cluster labeled as ``1," it is considered to have a high-expression neighborhood and is regarded as an overall SVG. However, MULTILAYER does not provide statistical significance for its overall SVG ranking.

\textbf{sepal} \cite{andersson2021sepal} is a diffusion-process-based method that identifies a gene as an overall SVG if its spatial expression pattern deviates from randomness. To quantify this deviation, sepal simulates a diffusion process from the gene's observed expression pattern until it converges to randomness, using the convergence time as a measure. The longer the convergence time, the more likely the gene is an overall SVG. However, this method ranks genes but does not provide statistical significance for the detected overall SVGs. Specifically, sepal uses the diffusion equation based on Fick's second law \cite{kantorovich1958approximate} to simulate the diffusion of the gene's transcripts in the 2D Euclidean space of a tissue slice. \rv{Note that sepal is included in the SRT analysis toolkit squidpy \cite{palla2022squidpy} for SVG detection.}

\textbf{BSP} \cite{wang2023dimension} is built upon the intuition that an overall SVG should have its spatial expression pattern vary significantly at different spatial resolutions; otherwise, the pattern should remain consistent. To implement this, BSP follows a four-step procedure. First, for each spot, it defines ``big" and ``small" patches consisting of neighboring spots within pre-specified large and small radii, respectively, so each spot has one big patch and one small patch. Second, for each gene, BSP calculates the average expression levels within the big patches and the variance of these averages, then similarly calculates the variance for the small patches. Third, BSP uses the ratio of the big-patch variance to the small-patch variance as a statistic to summarize the change in the gene's expression pattern with changing spatial resolution.  Finally, BSP ranks genes by this statistic, identifying those with large values as overall SVGs. Although BSP attempts frequentist inference by defining a null distribution for its statistic, this null distribution is improperly defined as a distribution fitted to the statistic values of all genes, implying that all genes are non-SVGs. \rv{Consequently, a p-value threshold of 0.05 will always lead to 5\% of genes being detected as overall SVGs, regardless of how many genes are truly overall SVGs.} Therefore, we do not consider BSP to provide valid statistical inference in our review. \rv{While the ranking of genes by the BSP statistic could still be informative, the lack of valid statistical inference means that we do not have a threshold on the statistic to reflect a target FDR.}

\textbf{PROST} \cite{liang2024prost} determines if a given gene is an overall SVG based on computer image segmentation. First, treating the gene's spatial expression in a tissue slice as an image, PROST applies image segmentation techniques to detect multiple foreground regions, where the gene is considered expressed, and one background region, where the gene is treated as unexpressed. Second, PROST defines two statistics: (1) a significance factor summarizing the overall elevation of the gene's expression from the background region to the foreground regions, and (2) a separability factor measuring the overall homogeneity of the gene's expression within each foreground region. Finally, PROST combines these two statistics into a single index. Based on this index, PROST ranks genes from high to low and selects the top-ranking genes as overall SVGs. Although PROST performs frequentist statistical tests, they are for Moran?s I \cite{moran1950notes} rather than its index. Therefore, we do not consider PROST to provide valid statistical inference in our review.

\rv{
\textbf{HEARTSVG} \cite{yuan2024heartsvg} is a time-series-based method designed to detect spatially variable genes (SVGs) by first converting a gene's expression levels in a 2D tissue slice, structured as a matrix with \( n_\text{row} \) rows and \( n_\text{col} \) columns, into four one-dimensional time series. Notably, HEARTSVG assumes that the tissue slice is square, which may impose a limitation. These four time series, termed ``marginal expression series," are generated through a process called ``semi-pooling." For example, if a tissue slice has \( n_\text{row} = 120 \) and \( n_\text{col} = 60 \), HEARTSVG defines the following four time series for each gene: (1) a per-row-average time series with 120 entries, each representing the average expression level within the corresponding row; (2) a within-row-window-average time series with \( 120 \times \frac{60}{\ln(120)} = 1440 \) entries, where each entry is derived from averaging expression levels within a non-overlapping sliding window of size \( \ln(120) = 5 \) within each row; (3) a per-column-average time series with 60 entries, each representing the average expression level within the corresponding column; and (4) a within-column-window-average time series with \( 60 \times \frac{120}{\ln(60)} = 1800 \) entries, where each entry is derived from averaging expression levels within a non-overlapping sliding window of size \( \ln(60) = 4 \) within each column. HEARTSVG then applies the Portmanteau test to each of these four time series to assess whether there are significant autocorrelations, which would suggest that the time series displays a trend or periodic pattern, potentially indicating an informative spatial pattern in the original 2D square. The four p-values, obtained from the Portmanteau tests on the four time series, are then combined into a single p-value using Stouffer?s method; that is, the gene should be considered an overall SVG if the combined p-value falls below a certain threshold. Finally, to determine the appropriate p-value threshold and account for multiple testing across all genes, HEARTSVG applies Holm's method to adjust the p-values and control the familywise error rate (FWER). It is important to note that HEARTSVG may have certain conceptual limitations, particularly the requirement for a square tissue slice and the seemingly arbitrary definitions of the four time series, especially regarding the chosen window sizes and the construction of the time series, given the many possible alternatives.
}

\textbf{SPADE} \cite{bae2021SPADE} leverages the hematoxylin-and-eosin (H\&E) image accompanying the SRT data (e.g., 10x Visium) to detect overall SVGs. It defines a gene as an overall SVG if the gene's expression levels can be predicted by H\&E features, which are extracted by a pre-trained convolutional neural network and expected to contain critical spatial information, in a linear model. Specifically, SPADE first performs principal component analysis (PCA) on the $512$ H\&E features for dimensionality reduction. Then, it tests if the resulting principal components can predict the gene's expression levels in a linear model. If so, the gene is defined as an overall SVG.

\textbf{BOOST-MI} \cite{jiang2022BOOSTMI} and \textbf{BOOST-HMI} \cite{yang2024bayesian} are Bayesian methods that detect overall SVGs based on the Ising model and its extension to the geostatistical mark interaction model, respectively. The Ising model, historically used to model ferromagnetism, represents each magnetic moment (or spin) in a material as a discrete variable that can take one of two values: +1 or -1 (interpreted as spin up or spin down). In ferromagnetic materials, spins tend to align with their neighbors to minimize the system's energy. The Ising model exhibits a phase transition at a critical temperature, below which spins align, resulting in net magnetization corresponding to the ferromagnetic phase. Above this temperature, thermal fluctuations dominate, and spins are randomly oriented, corresponding to the paramagnetic phase. BOOST-MI applies the Ising model to SRT data by treating spatial spots as spins, with the positive and negative values corresponding to high and low expression levels of a gene after dichotomization. BOOST-MI detects a gene as an overall SVG if it finds the gene to be in a "ferromagnetic phase." BOOST-HMI extends this approach for imaging-based SRT data to accommodate the irregular spatial distribution of measured spots.


 \subsection{Graph-based methods} 
In manifold learning, when data originate from a non-linear manifold within Euclidean space, using a graph to represent the data effectively captures the intrinsic geometry and preserves the local structures of the manifold. Extending this concept to spatial spots on a non-linear manifold within the 2D Euclidean space of a tissue slice, graph-based SVG detection methods first construct a neighborhood graph by connecting nearby spots in Euclidean space. These methods then operate on the graph to detect overall SVGs. The way the graph is constructed is crucial, as it determines how well the graph represents the manifold, significantly impacting the performance of graph-based methods.

In this section, we introduce nine graph-based methods (Fig.~\ref{fig:decisiontree}), focusing on their graph construction approaches and subsequent operations on the graph for detecting overall SVGs.

\textbf{Hotspot} \cite{detomaso2021hotspot} aims to detect overall SVGs as the genes that exhibit high expression levels in local spot neighborhoods (i.e., ?hotspots?). It first constructs a $K$-nearest neighbor (KNN) directed graph by treating each spot as a node and connecting it to its closest $K$ spots in Euclidean distance. For a spot $i$ connected to a spot $j$, with Euclidean distance denoted by $d_{ij}$, the directed edge from $i$ to $j$ is assigned a weight defined as \(w_{ij} = e^{-d_{ij}^2 / \sigma_i^2}\), where $\sigma_i$ represents the bandwidth of spot $i$ (defined as the distance from spot $i$ to its $[K/3]$-th neighbor). Next, for each gene, denoting its expression count at spot $i$ by $Y_i$, Hotspot uses an autocorrelation statistic \(H\) to quantify the dependence of the gene's expression level on the graph structure:
\[
H = \sum_i \sum_{j \neq i} w_{ij} Y_i Y_j\,,
\]
which is related to Moran's I \cite{moran1950notes}, a spatial autocorrelation measure that quantifies the degree to which similar expression levels are clustered together in space. Moran's I is defined as
     \begin{equation}\label{eq:moranI}
         I = \frac{n}{\sum_i \sum_j w_{ij}} \frac{\sum_i \sum_j w_{ij} (Y_i - \bar{Y})(Y_j - \bar{Y})}{\sum_i (Y_i - \bar{Y})^2}\,,
     \end{equation}
     where \(n\) is the number of spatial spots, and \(\bar{Y}\) is the gene's mean expression level across spots.
Similar to $I$, a large \(H\) indicates that spots where the gene is highly expressed are clustered in local neighborhoods, suggesting the gene is an overall SVG. To assess the statistical significance of \(H\), Hotspot converts \(H\) to a z-statistic by subtracting the expectation of \(H\) under the null model, where all spots are assumed to be independent, and dividing by the standard deviation of \(H\) under the null model. Specifically, Hotspot assumes two null models where \(Y_i\) independently follows either an NB distribution or a Bernoulli distribution. The z-statistic is then assumed to follow the standard Gaussian distribution under the null hypothesis, and a one-sided p-value is computed.

\textbf{HRG} \cite{wu2022highly} differs from Hotspot in three key aspects. First, it replaces the KNN graph with a shared-nearest-neighbors (SNN) graph, where the edge weight \( w_{ij} \) between spots \( i \) and \( j \) is defined as the Jaccard index of their respective \( K \)-nearest neighbors. The Jaccard index is calculated as the size of the intersection of the two sets divided by the size of their union. Second, HRG modifies \( Y_i \), the gene count at spot \( i \), to a z-score \( Z_i = \frac{Y_i - \bar{Y}}{\sqrt{\frac{1}{n}\sum_{j=1}^n(Y_j - \bar{Y})^2}} \), where \( \bar{Y} \) is the mean gene count and \( n \) is the total number of spots. Third, unlike Hotspot, HRG does not provide statistical significance for its detected overall SVGs but instead ranks the genes based on \( H \).

\textbf{SVGbit} \cite{hong2023spatiotemporal} is also a spatial-autocorrelation-based method that ranks genes as overall SVGs based on the spatial autocorrelations of local neighborhoods. The method does not provide statistical significance for its rankings. For each gene, SVGbit uses three steps to calculate a summary statistic, called the ``aggregation index," which is used to rank genes.  
First, SVGbit identifies ``hotspots" as spatial spots with large and statistically significant local Moran's I values, using a z-test-based p-value thresholded at the $5\%$ FDR. Specificallhy, for spot $i$, local Moran's I is defined as 
\[ I_i = (Y_i - \bar{Y}) \sum_{j} w_{ij} (Y_j - \bar{Y})\,, \]
where the terms are defined similarly to Moran's I \eqref{eq:moranI}. A high \( I_i \) value indicates that spot \( i \) is part of a neighborhood of similar values (either high-high or low-low).
Hence, spot $i$ is treated as a hotspot if $I_i$ is deemed significantly large. Second, for each hotspot, SVGbit considers its neighboring hotspots among the $6$-nearest-neighbor spots and defines the ``local aggregation density" as the sum of these neighboring hotspots' normalized log-transformed p-values (i.e., the log-transformed p-value of each neighboring hotspot divided by the sum of the log-transformed p-values of the $6$ neighboring spots). Thus, a hotspot with more neighboring hotspots has a higher local aggregation density. Third, the gene's aggregation index is computed as the average of all hotspots' local aggregation densities. Intuitively, the overall SVGs with large aggregation indices are those whose expression patterns exhibit clustered hotspots.


    \textbf{SINFONIA} \cite{jiang2023sinfonia} aims to detect overall SVGs as those exhibiting positive spatial autocorrelations, where similar expression levels are clustered in neighboring spots, either globally or locally. It begins by constructing a KNN directed graph, setting the edge weight from spot \(i\) to spot \(j\) (if connected) as \(w_{ij} = 1 - \frac{d_{ij}}{\max_{j} d_{ij}}\), where \(d_{ij}\) is the Euclidean distance between spots \(i\) and \(j\), and \(\max_{j} d_{ij}\) is the maximum distance between spot \(i\) and its \(K\) nearest neighbors. Given this graph, SINFONIA calculates two measures of spatial autocorrelation for each gene: (1) Moran?s I \cite{moran1950notes}, which provides a global measure of spatial autocorrelation, with values ranging from $-1$ to $1$, where a value close to $1$ indicates strong positive spatial autocorrelation; (2) Geary?s C \cite{geary1954contiguity}, which is more sensitive to local spatial differences, with values ranging from $0$ to $2$, where a value close to $0$ indicates strong positive spatial autocorrelation. Geary's C is defined as
    \[
     C = \frac{(n-1)}{2\sum_i \sum_j w_{ij}} \frac{\sum_i \sum_j w_{ij} (Y_i - Y_j)^2}{\sum_i (Y_i - \bar{Y})^2}\,,
     \]
     where the terms are defined similarly to Moran's I \eqref{eq:moranI}.
    SINFONIA then ranks genes from high to low based on Moran's I and from low to high based on Geary's C. To identify overall SVGs, SINFONIA takes the union of the top \(J\) genes from both rank lists, where \(J\) is a pre-specified positive integer. However, SINFONIA does not provide statistical significance for the detected overall SVGs.

    \textbf{MERINGUE} \cite{miller2021characterizing} also uses Moran's I to detect overall SVGs but differs from previous methods in its graph construction approach. Instead of using a KNN graph, MERINGUE constructs a graph of spatial spots using Delaunay triangulation, making it suited for spatial spots with non-uniform density, such as cells in MERFISH data. Delaunay triangulation \cite{preparata2012computational} connects points in a 2D space to form triangles such that no point lies inside the circumcircle (the circle that passes through all three vertices) of any triangle. In simpler terms, it maximizes the minimum angle of the triangles, avoiding skinny triangles. Unlike KNN, Delaunay triangulation does not require selecting an arbitrary \( K \) parameter and automatically adapts to variations in point density, providing more connections in denser regions and fewer in sparser regions. Given the graph, MERINGUE assigns an edge weight between spots \( i \) and \( j \) as \( w_{ij} = 1 \) if the two spots are connected, and \( w_{ij} = 0 \) otherwise. To determine if a gene is an overall SVG, MERINGUE uses Moran's I, transforms it to a z-statistic, and calculates a one-sided p-value based on the standard Gaussian distribution.
    
\textbf{SpaGene} \cite{liu2022scalable}, similar to previous methods that use Moran's I, aims to detect overall SVGs as those highly expressed at neighboring spots. Starting from a KNN graph of spots, it uses an alternative approach to boost computational efficiency. For each gene, it dichotomizes spots, coding them as 0 or 1 corresponding to low or high expression levels of the gene. It then extracts the high-expression spots (i.e., the spots labeled as 1) from the graph and summarizes the distribution of these spots' degrees (the number of edges connected to each spot) in the subgraph. Finally, it compares the degree distribution of these high-expression spots to that of the whole graph, and calculates the earth mover's distance between the two distributions. A small distance indicates that the high-expression spots are densely connected to each other and rarely connected to other spots, so the gene is likely an overall SVG. Finally, SpaGene converts the distance to a p-value using a permutation test.

    \textbf{BinSpect}, part of the Giotto toolbox \cite{dries2021giotto}, detects overall SVGs by examining whether the dichotomized spots (i.e., spots coded as 0 or 1 corresponding to low and high expression levels of a gene, respectively) are randomly distributed in a graph. Specifically, BinSpect constructs a graph of spots either as a KNN graph or by Delaunay triangulation. Then, for each gene, BinSpect follows three steps to determine if the gene is an overall SVG. First, it assigns a binary label to each spot by either thresholding at the top $30\%$ expressed spots or using $k$-means clustering ($k=2$). Second, it constructs a 2-by-2 contingency table by counting the edges that connect spots with labels (0,0), (0,1), (1,0), or (1,1). Finally, it applies Fisher's exact test to the contingency table to decide whether spots with the same label tend to be connected, thereby detecting a gene as an overall SVG.

\rv{Note that, unlike Giotto, two other popular toolkits with SRT functionalities---Seurat and squidpy---implement different methods for detecting overall SVGs. Seurat \cite{seurat} uses two approaches: Moran's I and a mark-point-process-based method called ``markvariogram" similar to Trendsceek \cite{edsgard2018trendsceek}. On the other hand, squidpy \cite{palla2022squidpy}  employs sepal \cite{andersson2021sepal}, a diffusion-process-based method.}

\textbf{scGCO} \cite{zhang2022identification} is similar to BinSpect at a high level by testing whether a gene's discretized expression levels are dependent on spatial neighborhoods. However, scGCO approaches the problem in a more complex way. After constructing an undirected graph of spots using Delaunay triangulation, scGCO performs three steps. First, it fits a Gaussian mixture model to the gene's log-transformed expression levels to discretize the gene's expression into a number of levels. Second, using a graph cut algorithm, it fits a hidden Markov random field to the gene's log-transformed expression levels on the spot graph to divide the spots into segments. Finally, it tests whether each discretized level is randomly distributed in each segment, using the homogeneous Poisson process as the underlying null model. In summary, a gene is considered an overall SVG if at least one of its discretized levels is non-randomly distributed in any segment.

  \textbf{RayleighSelection} \cite{govek2019clustering} is a theoretical approach that generalizes a graph to a simplicial complex \cite{horak2013spectra}, incorporating additional higher-dimensional topological information about spatial spots compared to a graph. The method first constructs the Vietoris-Rips complex, a type of simplicial complex, of spatial spots by considering balls of a fixed radius around each spot and forming edges, triangles, and higher-dimensional simplices based on the intersections of these balls. For each gene, it then computes the combinatorial Laplacian score, which measures how well the gene respects the local structure of the simplicial complex. If a gene has high expression levels in highly connected regions, it will have a low combinatorial Laplacian score, indicating a significant spatial pattern and suggesting it to be an overall SVG. Finally, RayleighSelection computes a p-value for the combinatorial Laplacian score using a permutation test.

\subsection{Relationship between Euclidean-space-based methods and graph-based methods}

When it is reasonable to assume data lie on a non-linear manifold in Euclidean space, using a graph to represent the data offers significant advantages. A graph can capture the intrinsic geometry of the manifold by representing data points as nodes and their similarity relationships as edges based on some similarity or distance measure. Graph-construction methods like KNN (where each node's neighborhood is defined by a fixed number of neighbors, $K$) or $\epsilon$-neighborhoods (where each node's neighborhood is defined by a fixed radius, $\epsilon$) ensure that each node is connected to its nearest neighbors, preserving the local relationships among nodes. The shortest-path distance between two nodes in the graph approximates their ?geodesic distance? on the manifold, reflecting the data?s intrinsic structure more accurately than the Euclidean distance. Additionally, graph-based methods are more robust to the curse of dimensionality: in high-dimensional spaces, where Euclidean distances often lose meaning, a graph can effectively capture relationships among data points.

However, in the context of SVG detection, using Euclidean distance to represent distances between spots might be more appropriate. Biological relationships between spots inherently follow Euclidean distances in a physical 2D Euclidean space in tissue. Moreover, since Euclidean distances are based on 2D spatial locations rather than high-dimensional gene expression data, the curse of dimensionality is not an issue.

It remains an open question how to construct a graph to accurately represent a non-linear manifold of spatial spots. The choice of graph-construction method, such as KNN or Delaunay triangulation, is crucial. KNN connects each spot to its $K$ nearest neighbors, preserving the local structure of the data. In contrast, Delaunay triangulation handles the non-uniform spatial distribution of spots in SRT data, such as MERFISH or Slide-seq data. Delaunay triangulation connects spots to form triangles such that no point is inside the circumcircle of any triangle, resulting in a graph that more naturally respects the non-uniform distribution of spots and preserves local density variations. However, Delaunay triangulation is more computationally expensive and more sensitive to outliers than KNN. \rv{Future studies could explore whether the computational cost of using Delaunay triangulation for imaging-based SRT data is justified, and how this decision may vary depending on the tissue structure?whether the tissue is structured (such as in the brain) or unstructured (such as in tumors). Additionally, research could focus on improving the scalability and robustness of Delaunay triangulation in these contexts.} 

Additionally, the definition of edge weights in the graph is another important consideration. Edges can be directional or undirectional, depending on whether the relationship between spots has a direction (e.g., one cell type influencing another cell type's gene expression). The edge weights can take multiple forms. For example, weights can be determined by a Gaussian kernel, where weights decay exponentially with the Euclidean distance between spots, emphasizing closer spots. Alternatively, weights can decrease linearly with distance. These choices affect how the graph captures the manifold structure of data and, consequently, how effectively it can be used to detect spatial patterns of gene expression that are biologically relevant.



\subsection{Applications of overall SVGs}
The detection of overall SVGs mainly serves as a pre-processing step in SRT data analysis. While the definition of overall SVGs may vary across methods, the common objective is to select informative genes for downstream analysis. 
A common downstream analysis is to identify \textit{spatial domains} (also referred to as \textit{spatial communities}) by partitioning a tissue slice into regions so that spots have similar expression profiles of overall SVGs in each region. 

Identifying spatial domains can help uncover tissue layers where morphological architecture is less defined. For example, SPADE \cite{bae2021SPADE}, an overall SVG detection method, uses its detected overall SVGs to identify spatial domains as substructures in cortical layers and amygdala in a 10x Visium mouse brain immunofluorescence dataset. Another use of spatial domain identification is to reveal gene expression profiles underlying tissue structures. For example, scGCO \cite{zhang2022scGCO}, another overall SVG detection method, identifies spatial domains from a mouse breast cancer biopsy dataset sequenced by Spatial Transcriptomics \cite{staahl2016visualization}, and the identified domains align with annotated tissue structures including invasive ductal cancer, ductal cancer, and normal tissue. This alignment confirms that these tissue structures have distinct gene expression profiles, so the SRT data can potentially lead to the discovery of new marker genes for these tissue structures, a task tackled by spatial-domain-marker SVG detection.

A common approach to identifying spatial domains is clustering spatial spots using overall SVGs' expression levels and spot locations. For example, graph-based clustering (e.g., Louvain clustering) can be applied after spatial spots are connected into a graph based on their spatial proximity and gene expression levels; then, the identified clusters contain spatial spots in proximity and exhibiting similar gene expression levels. For example, using Louvain clustering, the overall SVGs detected by SPADE \cite{bae2021SPADE} and SINFONIA \cite{jiang2023sinfonia} resulted in spatial domains in well-structured brain tissues in the mouse olfactory bulb Spatial Transcriptomics dataset \cite{staahl2016visualization}, as well as in the mouse hippocampus seqFISH \cite{shah2016situ} and Slide-seq \cite{stickels2021highly} datasets.

The spatial domains identified from overall SVGs can be validated, directly or indirectly, in three ways. First, spatial domains can be compared with tissue layers annotated by pathologists from the H\&E image accompanying the SRT data (e.g., 10x Visium). For example, in the SpaGCN \cite{hu2021spagcn} method paper, three annotated layers of the primary pancreatic cancer were compared to the spatial domains identified from a Spatial Transcriptomics dataset \cite{guo2022integrating}. Second, spatial domains can be annotated with cell-type labels in external transcriptomic data to verify if distinct domains have different cell-type compositions. For example, in a study of the human dorsolateral prefrontal cortex 10x Visium dataset \cite{maynard2021transcriptome}, researchers annotated the identified spatial domains as cortical layers based on the cytoarchitecture and marker genes obtained from external large-scale single nucleus RNA-seq datasets. Third, for well-structured tissue types like the brain, spatial domains can be verified by transferring tissue layer labels from external public annotations. For instance, in the SINFONIA \cite{jiang2023sinfonia} method paper, annotations from Allen Brain Atlas \cite{sunkin2012allen} and Mouse Brain Gene Expression Atlas (http://mousebrain.org/) were transferred to several SRT datasets, including the mouse brain coronal section 10x Visium dataset, the mouse hippocampus Slide-seqV2 dataset \cite{stickels2021highly}, and the mouse olfactory bulb STEREO-seq dataset \cite{chen2022spatiotemporal}. 

\rv{Although the detection of overall SVGs may serve as a feature screening step before spatial domain identification to circumvent the curse of dimensionality (i.e., dealing with too many genes as features), it is important to note that this screening step may not be necessary for spatial domain detection methods that employ alternative strategies to reduce feature dimensions. For example, BayesSpace \cite{zhao2021spatial} and StLearn \cite{pham2020stlearn} implement spatial domain identification using lower-dimensional representations of spatial spots based on selected HVGs.}

\rv{Besides spatial domain identification, another downstream analysis is the unsupervised identification of spatial gene modules among the detected overall SVGs. Each module is a cluster of overall SVGs with similar spatial expression patterns, representing a common spatial pattern shared by these genes (also referred to as a co-expression pattern). Compared to the spatial expression patterns of individual overall SVGs, the patterns of gene modules are less noisy and can provide clearer insights into the molecular architecture underlying tissue structures---specifically, which genes co-determine a tissue structure in a synchronized manner. For example, in downstream applications of MULTILAYER \cite{moehlin2021multilayer}, MERINGUE \cite{miller2021MERINGUE}, and Binspect in Giotto \cite{dries2021giotto}, the detected overall SVGs were grouped into modules using hierarchical clustering based on a constructed gene co-expression matrix or a gene-gene similarity graph.} 

\rv{It is worth noting that the identification of spatial gene modules can precede and facilitate the discovery of non-global spatial domains that are specific to one or more gene modules. For example, among the overall SVGs detected by MULTILAYER \cite{moehlin2021multilayer}, the genes \textit{ACTA2} and \textit{ELN} were found to be spatially co-expressed, leading to the identification of five tissue substructures in human heart tissue. Similarly, spatial gene modules derived from the overall SVGs detected by MERINGUE \cite{miller2021MERINGUE} enabled the detection of functional spatial regions in mouse brain tissue. Additionally, spatial gene modules identified from the overall SVGs detected by Binspect in Giotto \cite{dries2021giotto} successfully recovered known anatomical structures in mouse kidney tissue.}
\section{Methods for detecting cell-type-specific SVGs}
In both sequencing- and imaging-based SRT data, a gene's expression variance across spots can result from the gene's different expression levels in distinct cell types. Given that cell types are typically distributed non-uniformly within a tissue slice, neglecting the cell type information of spatial spots may hinder the discovery of genes whose expression patterns reflect informative spatial variation beyond the variation attributed to cell-type composition \cite{yu2022identification, cable2022robust}. \rv{Thus, we define \textit{cell-type-specific SVGs} as the genes that exhibit non-random spatial expression patterns within a cell type. These genes are detected using both SRT data and external cell-type annotations for the spatial spots (See Fig.~\ref{fig:catecartoon}a and Table~\ref{tab:concepts_summary}).} Unlike the methods for detecting overall SVGs that do not consider cell type annotations, the three methods for detecting cell-type-specific SVGs---\textbf{CTSV} \cite{yu2022identification}, \textbf{C-SIDE} \cite{cable2022cell}, and \textbf{spVC} \cite{yu2024spvc}---begin by annotating the cell types of the spatial spots in SRT data and then use a regression framework to identify SVGs within cell types by examining the interaction effects between cell types and spatial locations (see Section~\ref{subsec:regfixeffect} for a detailed discussion on the general framework).



\textbf{CTSV} assumes a ZINB distribution for a given gene's expression count \(Y_{i}\) at spot \(i\):
\[
Y_{i} \overset{\mathrm{ind}}\sim \operatorname{ZINB}(\mu_{i}(\mathbf{s}_i), \bm{\theta})\,,
\]
where \(\mu_{i}(\mathbf{s}_i)\) is the mean expression function of the spatial location \(\mathbf{s}_i = (s_{i1}, s_{i2})^\top\), and \(\bm{\theta}\) indicates the other parameters (dispersion and zero-inflated probability) necessary to describe the distribution. To determine if the gene is a cell-type-specific SVG, CTSV includes \(K\) cell types' cell-type-level mean functions of \(\mathbf{s}_i\): \(\eta_{k}(\mathbf{s}_i),~k=1, \ldots, K\), and assumes that 
\begin{align*}
\log \mu_{i}(\mathbf{s}_i) &= \log \ell_i  + \sum_{k=1}^K \eta_{k}(\mathbf{s}_i) w_{ik} \,,\\
\eta_{k}(\mathbf{s}_i) &= \beta_{k0} + \beta_{k1} b_1(s_{i1}) + \beta_{k2} b_2(s_{i2})\,.
\end{align*}
where $\ell_i$ denotes spot $i$'s library size (i.e., total expression count), and the weights \(w_{ik},~k=1, \ldots, K\), are pre-estimated cell type proportions of spot \(i\) obtained by spatial deconvolution methods such as SPOTlight \cite{elosua2021spotlight} and RCTD \cite{cable2022robust}, and \(b_1(s_{i1})\) and \(b_2(s_{i2})\) are two non-cell-type-specific, non-parametric, univariate functions of the spatial coordinates in the two dimensions. CTSV then tests if \(\beta_{k1}\) and \(\beta_{k2}\) are both zero. If not, the gene is considered an SVG specific to cell type \(k\).

\textbf{C-SIDE} is similar to CTSV but has three major differences. First, C-SIDE assumes a Poisson distribution for $Y_i$. Second, C-SIDE assumes a different relationship between $\mu_i(\mathbf{s}_i)$ and \(\eta_{k}(\mathbf{s}_i),~k=1, \ldots, K\). Third, C-SIDE uses a different non-parametric function form for each $\eta_{k}(\mathbf{s}_i)$.
\begin{align*}
Y_{i}  &\overset{\mathrm{ind}}\sim \operatorname{Poisson}\left(\mu_{i}(\mathbf{s}_i)\right)\,,\\
\log \left(\mu_{i}(\mathbf{s}_i) \right) &= \gamma_0 + \log\ell_i + \log \left(\sum_{k=1}^K \eta_{k}(\mathbf{s}_i) w_{ik}\right) + \epsilon_{i}\,,\\
\log \left(\eta_{k}(\mathbf{s}_i)\right) &= \beta_{k0} + \sum_{\ell=1}^L \beta_{k\ell} b_{\ell}(\mathbf{s}_i)\,,
\end{align*}
where the common terms are defined similarly to CTSV, $\epsilon_i$ is a zero-mean random-effect term, and $\sum_{\ell=1}^L \beta_{k\ell} b_{\ell}(\mathbf{s}_i)$ is a non-cell-type-specific, bivariate smooth-spline function of $\mathbf{s}_i$ consisting of $L$ basis functions. C-SIDE then tests if $\beta_{k1},\ldots,\beta_{kL}$ are all zero. If not, the gene is considered an SVG specific to cell type $k$.

\textbf{spVC} is similar to C-SIDE but assumes a different relationship between $\mu_i(\mathbf{s}_i)$ and \(\eta_{k}(\mathbf{s}_i),~k=1, \ldots, K\).
\begin{align*}
\log \left(\mu_{i}(\mathbf{s}_i) \right) &= \gamma_0 + \log\ell_i + \sum_{k=1}^K w_{ik} \beta_k + \sum_{k=1}^K \eta_k(\mathbf{s}_i) w_{ik} + \gamma(\mathbf{s}_i) \,,
\end{align*}
which considers cell types' non-spatial effects as $\sum_{k=1}^K w_{ik} \beta_k$ and a baseline spatial effect $\gamma(\mathbf{s}_i)$. Then, spVC implements a two-step testing procedure. First, it tests whether $\beta_1,\ldots,\beta_K$ are all zero and whether $\gamma(\cdot) = 0$ in a reduced model without the cell-type-specific spatial effect term $\sum_{k=1}^K \eta_k(\mathbf{s}_i) w_{ik}$. If both null hypotheses are rejected, the second step is to test whether $\eta_k(\cdot)=0$. If not, the gene is considered a SVG specific to cell type $k$.

\subsection{Applications of cell-type-specific SVGs}

\rv{Although the publications of the three cell-type-specific SVG detection methods \cite{yu2022identification, cable2022cell, yu2024spvc} did not specifically outline applications for cell-type-specific SVGs, we envision potential applications that extend the two major applications of overall SVGs---identifying spatial domains and spatial gene modules---to a more granular, cell-type level. First, cell-type-specific SVGs can be used to identify and characterize distinct cell subpopulations or states within particular cell types. Second, cell-type-specific SVGs can enable the identification of gene modules within specific cell types. These applications have the potential to provide new molecular insights into tissues with complex structures and diverse cell types, such as cancer and brain tissue. For example, in cancer research, cell-type-specific SVGs can help identify distinct states of cancer cells and immune cells, leading to a better understanding of tumor heterogeneity and the tumor microenvironment. Additionally, cell-type-specific SVGs can facilitate the identification of cell-type-specific gene modules that could be crucial for developing targeted therapies and personalized treatment strategies. Similarly, in brain tissue, cell-type-specific SVGs can identify diverse neuronal subpopulations that may correspond to functional brain regions. Furthermore, cell-type-specific SVGs can enable the discovery of cell-type-specific gene modules that may regulate neuronal activity and connectivity, potentially advancing our understanding of brain development and neurodegenerative diseases.} 

\section{Methods for detecting spatial-domain-marker SVGs}\label{sec
}

\rv{We define \textit{spatial-domain-marker SVGs} as the genes that exhibit significantly higher expression in a spatial domain compared to other domains. These genes are detected using SRT data and spatial domains, which are usually detected from the same SRT data (See Fig.~\ref{fig:catecartoon}a and Table~\ref{tab:concepts_summary}).} Unlike previous SVG detection methods, the two methods for detecting spatial-domain-marker SVGs?\textbf{SpaGCN} \cite{hu2021spagcn} and \textbf{DESpace} \cite{cai2024despace}?first partition spatial spots into more than one spatial domain. Then, they implement hypothesis tests to assess a gene's mean expression differences between these spatial domains. A gene is defined as a marker SVG of a spatial domain if \rv{shows significantly higher expression} in that domain than in other domains.

\textbf{SpaGCN} identifies spatial domains using a pre-trained graph convolutional network applied to SRT data and the paired H\&E image. For each gene, it performs Wilcoxon rank-sum tests on normalized expression levels between each domain and the neighboring spots. If the gene is found to  \rv{have significantly higher expression} in a domain, it is considered a marker SVG of that domain.

\textbf{DESpace} first implements existing spatial clustering methods, such as BayesSpace \cite{zhao2021spatial} and StLearn \cite{pham2023robust}, to SRT data to identify spatial clusters as spatial domains. \rv{DESpace then offers two modes for detecting SVGs: the first mode identifies spatial-domain-marker SVGs, while the second mode detects overall SVGs. In the first mode, DESpace} uses an NB generalized linear model to assess if the spatial domains have a significant effect on a gene's expression. If a significant effect is found, the gene is identified as a spatial-domain-marker SVG, associated with the domain where it has significantly higher expression compared to other domains. \rv{In the second mode, DESpace tests the null hypothesis that all spatial domains have the same effect on a gene's expression. If this hypothesis is rejected, indicating that the gene exhibits differences among spatial domains, the gene is detected as an overall SVG. Therefore, the spatial-domain-marker SVGs detected by DESpace must be included among the overall SVGs it identifies. Conversely, the overall SVGs detected by DESpace must serve as marker genes for some spatial domains.}

\subsection{Applications of spatial-domain-marker SVGs}
\rv{Spatial-domain-marker SVGs have two applications for characterizing predefined spatial domains in a tissue slice, particularly when these domains align with tissue structures. First, spatial-domain-marker SVGs can help elucidate the underlying molecular mechanisms of tissue structures. For example, SpaGCN \cite{hu2021spagcn} detected spatial-domain-marker SVGs \textit{KRT17} and \textit{MMP11} in a pancreatic cancer region, which align well with pancreatic cancer biology. \textit{KRT17} functions as a tumor promoter and regulates proliferation in pancreatic cancer, while \textit{MMP11} is a prognostic biomarker for the disease. Second, spatial-domain-marker SVGs can assist in annotating tissue structures in other SRT datasets. For instance, the spatial-domain-marker SVGs identified by SpaGCN \cite{hu2021spagcn} from a human dorsolateral prefrontal cortex slice (No. 151673 in the LIBD (Lieber Institute for Brain Development) SRT dataset) were used to annotate the tissue layers in a different brain slice (No. 151507).}

\section{Theoretical characterization of SVG detection methods that use frequentist hypothesis tests}\label{sec:test}

Among the 33 SVG detection methods, 23 of them
implement statistical hypothesis tests to detect SVGs using \textit{frequentist inference} (i.e., defining a test statistic, deriving the test statistic's null distribution, and converting the test statistic value to a p-value). For these fequentist hypothesis-testing-based methods, their null and alternative hypotheses direct their SVG detection. In general, methods that use different null hypotheses are not directly comparable. The reason is that a gene may satisfy one null hypothesis (and is considered a true non-SVG) but not the other null hypothesis (and is defined as a true SVG). Hence, we would like to clarify and categorize the null hypotheses used in the 23 methods to deepen our understanding of these methods' conceptual similarities and differences. 

\begin{figure}[htbp]
    \centering
    \includegraphics[width=\textwidth]{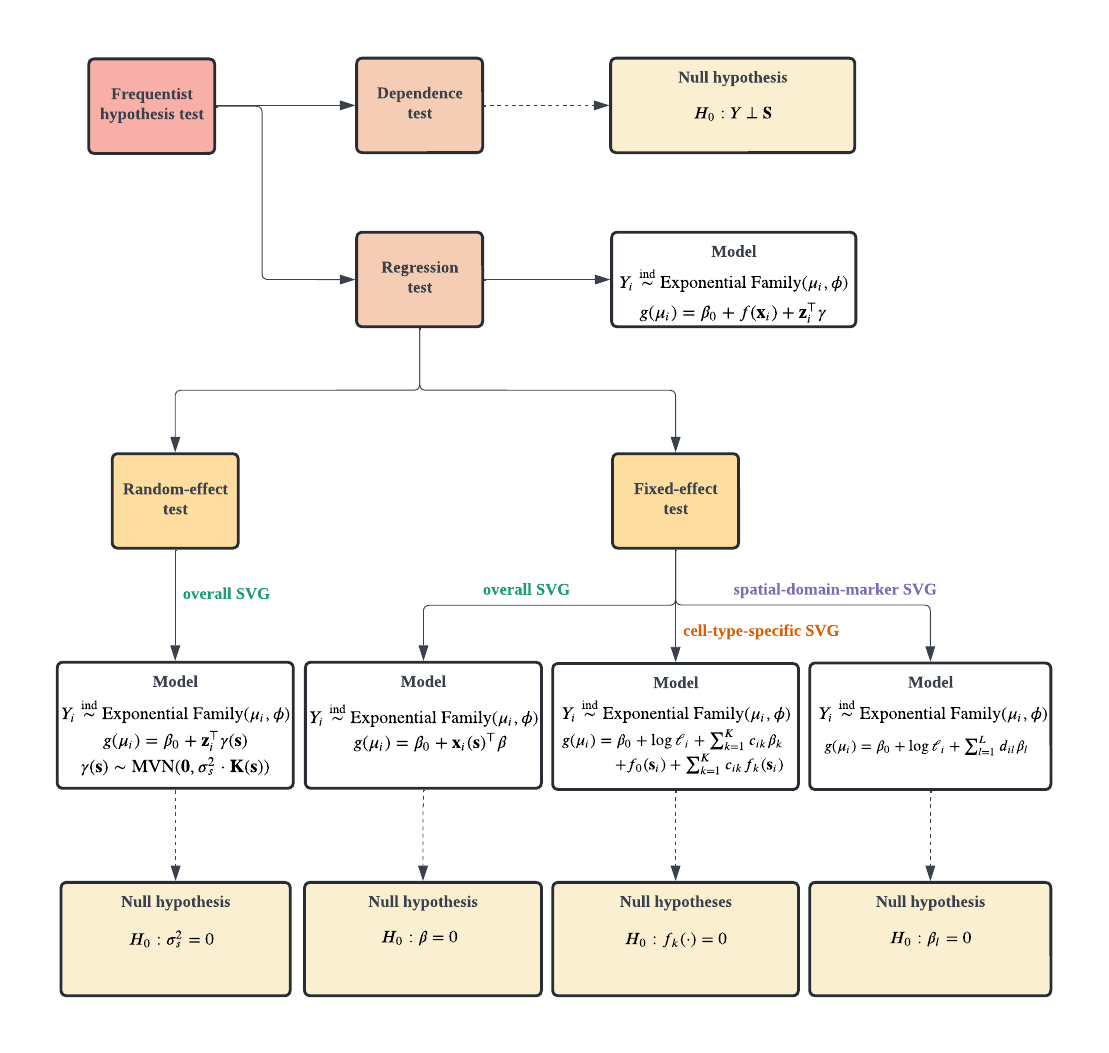}
    \caption{
    \textbf{Conceptual diagram for Section ``Theoretical characterization of SVG detection methods that use frequentist hypothesis tests."} This diagram illustrates the logical relationships among the three types of statistical tests (dependence test, regression fixed-effect test, and regression random-effect test) used by the 23 SVG detection methods that rely on frequentist hypothesis tests. The diagram also introduces the general form of statistical models upon which regression-based tests are performed and the corresponding null hypotheses for detecting SVGs.
    }
    \label{fig:conceptdiagram}
\end{figure}

Based on the types of null hypotheses, in Table~\ref{tab:cate} we summarize the hypothesis tests used in the 23 methods into three types: dependence tests, regression fixed-effect tests, and regression random-effect tests (also known as variance component tests). So far, dependence tests and regression random-effect tests have only been used for overall SVG detection, while regression fixed-effect tests have been used to detect all three categories of SVGs. For each method, besides the test type, we also list the test statistic and null distribution in Table~\ref{tab:tab_test}. \rv{Moreover, we define the three test types in Table~\ref{tab:concepts_summary} and include a conceptual diagram in Fig.~\ref{fig:conceptdiagram} that illustrates the relationships between the three test types and the three SVG categories.}

Among the three types of hypothesis tests for SVG detection \rv{(Fig.~\ref{fig:conceptdiagram})}, dependence tests have the most general null hypothesis: a gene's expression level is independent of the spatial location. In contrast, the two types of regression tests rely on specific assumptions for a regression model, which has a gene's expression level as the response variable and the spatial location as the predictor (also known as the covariate or explanatory variable). The co-existence of dependence tests and model-specific regression tests is common in the statistics literature \cite{li2024dissecting}. In general, dependence tests, thanks to their general independence null hypothesis, can capture SVGs with more diverse patterns but can be less powerful for detecting SVGs of specific patterns, compared to the regression tests. In contrast, relying on specific model assumptions, regression tests are more powerful for discovering the SVGs satisfying the assumptions but, meanwhile, more prone to false discoveries when the model assumptions do not hold \cite{li2022exaggerated}. Moreover, an advantage of regression tests is that they can more easily incorporate other covariates for adjustment compared to dependence tests. Additionally, their test statistics typically have theoretical null distributions, making p-value calculation straightforward and efficient. It remains an open question to benchmark the three types of tests regarding robustness to model misspecification and the trade-off between robustness and power. Additionally, it is important to identify which assumptions are more reasonable for SRT data generated from various tissues and by different technologies.





\subsection{Dependence tests}
For a given gene, the most general hypothesis test for SVG detection is to decide whether the gene's expression level $Y$ is independent of the spatial location $\mathbf{S}$, i.e., the null hypothesis is 
\begin{eqnarray*}
    H_0: Y \perp \mathbf{S}.
\end{eqnarray*}
In this formulation \rv{(Fig.~\ref{fig:conceptdiagram})}, we assume that $(y_1, \mathbf{s}_1), \ldots, (y_n, \mathbf{s}_n)$  are independently sampled from the joint distribution of $(Y, \mathbf{S})$, where $(y_i, \mathbf{s}_i)$ indicates spot $i$'s expression level of the given gene and spatial location. When the independent null hypothesis $H_0$ does not hold, the gene is a \textit{true overall SVG}.

Out of the 23 frequentist hypothesis-testing-based SVG detection methods, 11 methods adopt the dependent test formulation, constructing a test statistic to summarize the dependence between a gene's expression level and the spatial location. \rv{The 11 methods are SPARK-X, SMASH, singlecellHaystack, Trendsceek, HEARTSVG, Hotspot, MERINGUE, SpaGene, BinSpect, scGCO, and RayleighSelection. We categorize the test statistics used in the 11 methods into two types: \rv{test statistics with theoretical null distributions} (SPARK-X, SMASH, Hotspot, HEARTSVG, MERINGUE, BinSpect, and scGCO) and \rv{test statistics without theoretical null distributions} (Trendsceek, singlecellHaystack, RayleighSelection, and SpaGene).}

Among the \rv{seven} methods that use \rv{test statistics with theoretical null distributions}, SPARK-X is a representative method \cite{zhu2021spark}. The SPARK-X test statistic is defined to capture the ``agreement" between two spot similarity matrices (referred to as ``covariance matrices" by the SPARK-X authors), one defined based on a gene's expression levels at the $n$ spots and the other based on the spatial locations of the $n$ spots (subject to a transformation of the locations before defining the similarity between spots, so that the similarity matrix reflects a specific spatial pattern such as the Gaussian or cosine kernel). Specifically, the test statistic is defined as \((1/n)\) times the trace of the product of the two similarity matrices, which is essentially the Pearson correlation between the two vectorized matrices. The rationale is that under the independence null hypothesis, the test statistic should be small and follow a theoretical, asymptotic null distribution as a mixture chi-square distribution.

Four methods \rv{use test statistics that lack theoretical null distributions and therefore employ a permutation strategy.} They generate the null distribution of a test statistic by permuting a gene's expression levels across the $n$ spots, thus removing any dependence between the gene's expression level and the spatial location. The permutation procedure is a general solution for obtaining the null distribution when the theoretical null distribution is difficult to derive, but it is computationally intensive due to the need for repetitive permutations to generate the null distribution. For example, singlecellHaystack \cite{vandenbon2020singlecellHaystack} defines a test statistic based on the KL divergence of the conditional density of spots' spatial locations, which is conditional on a gene's dichotomized expression level (on or off), from the unconditional density of spots' spatial locations. Specifically, both densities are defined for each spatial grid, which consists of many spots and is pre-defined by the singlecellHaystack algorithm. Then, the test statistic is a summation of the KL-divergence-like statistics across the spatial grids. As this test statistic is complex and has no theoretical null distribution, the permutation is used to generate the null distribution.

\subsection{Regression tests}\label{sec:reg_test}
There are two types of regression tests \rv{(Fig.~\ref{fig:conceptdiagram})}: fixed-effect tests, where the effect of the spatial location is assumed to be fixed, and random-effect tests, which assume the effect of the spatial location as random. To explain these two types of tests, we start with a \textit{linear mixed model} for a given gene
\begin{eqnarray}\label{mod:lm}
    Y_i = \beta_0 + \mathbf{x}_i^{\top}\bm{\beta} + \mathbf{z}_i^{\top}\bm{\gamma} + \epsilon_i\,,
\end{eqnarray} 
where the response variable $Y_i$ is the gene's expression level at spot $i$, $\mathbf{x}_i \in \mathbb{R}^{p}$ indicates the fixed-effect covariates of spot $i$, $\mathbf{z}_i \in \mathbb{R}^{q}$ denotes the random-effect covariates of spot $i$, and $\epsilon_i$ is the random measurement error at spot $i$ with $\mathbb{E}[\epsilon_i]=0$. In the model parameters, $\beta_0$ is the (fixed) intercept, $\bm{\beta} \in \mathbb{R}^{p}$ indicates the fixed effects, and $\bm{\gamma}\in \mathbb{R}^{q}$ denotes the random effects with zero means $\E[\bm{\gamma}]=0$ and the covariance matrix $\Var(\bm{\gamma})=\bm{\Sigma} \in \mathbb{R}^{q \times q}$. In this linear mixed model, independence is assumed between $\bm{\gamma}$ and $\bm{\epsilon}=(\epsilon_1,\ldots,\epsilon_n)^{\top}$ and among $\epsilon_1,\ldots,\epsilon_n$. 

Fixed-effect tests examine whether some or all of the fixed-effect covariates $\mathbf{x}_i$ contribute to the mean of the response variable $\E[Y_i]$. If all fixed-effect covariates make no contribution, then $\E[Y_i | \mathbf{x}_i] = \E[Y_i]$. The null hypothesis 
\[H_0:\bm{\beta} = \mathbf{0}\,,
\]
implies $\E[Y_i | \mathbf{x}_i] = \E[Y_i]$, $i=1,\ldots,n$.

Random-effect tests examine whether the random-effect covariates contribute to the variance of the response variable $\Var(Y_i)$, focusing on the decomposition 
\begin{eqnarray} \label{eqn:ref_randomeff}
\Var(Y_i) = \Var(\E[Y_i | \mathbf{z}_i]) + \E[\Var(Y_i|\mathbf{z}_i)] = \mathbf{z}_i^{\top} \bm{\Sigma} \mathbf{z}_i + \Var(\epsilon_i)
\end{eqnarray} 
and testing if the contribution of the random-effect covariates $\Var(\E[Y_i | \mathbf{z}_i])$ is zero. The null hypothesis 
\[ H_0:\bm{\Sigma} = \mathbf{0}
\]
implies $\Var(\E[Y_i | \mathbf{z}_i]) = 0$, $i=1,\ldots,n$.

To generalize the linear mixed model \eqref{mod:lm} for data where the response variable $Y_i$ is not continuous (e.g., binary or count), we can decompose the model into a \textit{random structure}, where each $Y_i$ independently follows a distribution with mean $\mu_i = \mathbb{E}[Y_i]$, and a \textit{systematic structure} that links $\mu_i$ to the fixed-effect covariates $\mathbf{x}_i$. Specifically, for the model \eqref{mod:lm}, if we assume that $\bm{\gamma} \sim \operatorname{MVN}(\mathbf{0}, \bm{\Sigma})$ and $\epsilon_i \sim N(0, \sigma^2)$, then the \textit{random structure} is 
\[Y_i \overset{\mathrm{ind}}\sim N(\mu_i, \,\mathbf{z}_i^{\top} \bm{\Sigma} \mathbf{z}_i + \sigma^2)\,,
\]
and the \textit{systematic structure} is 
\[\mu_i = \beta_0 + \mathbf{x}_i^{\top}\bm{\beta}\,.
\]
The generalization of the linear mixed model can occur in both the random and systematic structures. For simplicity, we only focus on the fixed-effect covariates $\mathbf{x}_i$ for the generalization and omit the random-effect covariates $\mathbf{z}_i$. If we change the random structure from the Gaussian distribution to another exponential-family distribution, we have
\begin{align}\label{mod:rs}
    Y_i \overset{\mathrm{ind}}\sim \operatorname{Exponential~Family}(\mu_i, \phi)\,,
\end{align}
where $\phi$ is a nuisance parameter not of primary interest. An example of exponential-family distribution is the Poisson distribution $Y_i \sim \operatorname{Poisson}(\mu_i)$.  According to the random structure change, the systematic structure can be written in general as 
\begin{align}\label{mod:glm-ss}
    g(\mu_i) = \beta_0 + \mathbf{x}_i^{\top}\bm{\beta}\,,
\end{align}
where the function $g(\cdot)$ is referred to as the \textit{link function} and specified based on the distribution in the random structure. We refer to the model specified by \eqref{mod:rs}--\eqref{mod:glm-ss} as a \textit{generalized linear model}. If we further generalize the systematic structure so that the effects of the $p$ covariates in $\mathbf{x}_i = (x_{i1},\ldots,x_{ip})^{\top}$ on $\mu_i$ is non-linear and additive: 
\begin{align}\label{mod:gam-ss}
    g(\mu_i) = \beta_0 + \sum_{j=1}^p f_j(x_{ij})\,,
\end{align}
we referred to the model specified by \eqref{mod:rs} and \eqref{mod:gam-ss} as a \textit{generalized additive model}.

For regression models with different complexities (roughly speaking, different numbers of parameters to estimate), model selection centers on the bias-variance trade-off. An oversimplified model misses key data characteristics, leading to biased parameter estimates, though with low variance. In contrast, an overly complex model wastes parameters on unimportant details, increasing variance in parameter estimates. As a trade-off, a reasonable model appropriately addresses the data characteristics without overfitting the data noise, thereby being more powerful and robust.

For SVG detection, the effect of the spatial location $\mathbf{s}_i$ on the given gene's expression level $Y_i$ can be formulated as either fixed or random. In the following two subsections, we will discuss the SVG detection methods that adopt each formulation.

\subsubsection{Regression fixed-effect tests}\label{subsec:regfixeffect}

 Among the 23 frequentist hypothesis-testing-based SVG detection methods, six methods use the regression fixed-effect test formulation:  SPADE, C-SIDE, CTSV, spVC, SpaGCN, and DESpace. Notably, unlike dependence tests and regression random-effect tests, only regression fixed-effect tests cover all three SVG categories: overall SVGs, cell-type-specific SVGs, and spatial-domain-marker SVGs. Methods within each category use the same fixed-effect covariates $\mathbf{x}_i$ for spot $i$, while methods of different categories use different $\mathbf{x}_i$'s. Below, we introduce the six methods in the three categories. 

Among the six methods, SPADE \cite{bae2021SPADE} is the only method that detects overall SVGs. For a given gene, it uses a linear model without random-effect covariates:
	\begin{align*}
		\mu_i = \beta_0 + \mathbf{x}_i(\mathbf{s})^{\top}\bm{\beta}\,,
	\end{align*}
 where the fixed-effect covariates $\mathbf{x}_i(\mathbf{s})$ are defined as some processed features from the $n$ spots' spatial locations $\mathbf{s}$ and an H\&E image. Specifically, the processed features are based on $512$ features from a pre-trained convolutional neural network applied to an H\&E image. The gene is a \textit{true overall SVG} if $\bm{\beta} \neq \mathbf{0}$. SPADE uses the R package limma \cite{ritchie2015limma} to fit the linear model and tests if each component of $\bm{\beta}$ is zero using a t-test.

For cell-type-specific SVG detection, the fixed-effect covariates of spot $i$ (i.e., $\mathbf{x}_i$) include the library size $\ell_i$ (i.e., spot $i$'s total expression count), the cell type $\mathbf{c}_i = (c_{i1},\ldots,c_{iK})^{\top} \in [0,1]^K$, and the spatial location $\mathbf{s}_i \in \mathbb{R}^2$ (see Section~\ref{sec:intro} for definitions). The considered fixed effects include the marginal effects of $\mathbf{c}_i$ and $\mathbf{s}_i$ and their interactive effects. Hence, in a generalized additive model formulation for a given gene, the systematic structure is
 \begin{eqnarray}\label{mod:celltypespecific_fixed}
      g(\mu_i) = \beta_0 + \log\ell_i + \sum_{k=1}^{K} c_{ik}\beta_{k} + f_0(\mathbf{s}_i) + \sum_{k=1}^{K} c_{ik}f_{k}(\mathbf{s}_i),
 \end{eqnarray}
 where $\beta_0$ is the overall intercept, $\log \ell_i$ is the intercept effect of spot $i$'s library size, $\beta_k$ indicates the cell-type-specific spatial-invariant effect for cell type $k$, $f_0(\mathbf{s}_i)$ is the overall spatial effect at spot $i$, and $f_k(\mathbf{s}_i)$ is the cell-type-specific spatial effect for cell type $k$ at spot $i$. 
For identifiability, constraints are needed for $\bm{\beta}=(\beta_1, \ldots, \beta_K)^{\top}$ and $f_0(\cdot)$, $ f_1(\cdot), \ldots, f_K(\cdot)$. The gene is a \textit{true cell-type-specific SVG} of cell type $k$ if $f_k(\cdot) \neq 0$.

C-SIDE, CTSV, and spCV are three methods for cell-type-specific SVG detection. As an example method, spVC \cite{yu2024spvc} assumes a gene's expression count $Y_i$ follows a Poisson distribution (i.e., the random structure \eqref{mod:rs}) with the systematic structure (\ref{mod:celltypespecific_fixed}). To decide whether the gene is a cell-type-specific SVG, spVC implements a two-step procedure for sequential hypothesis tests. First, it considers a reduced model without interactive effects between $\mathbf{c}_i$ and $\mathbf{s}_i$, so the systematic structure becomes
 \begin{eqnarray*}
      g(\mu_i) = \beta_0 + \log\ell_i + \sum_{k=1}^{K} c_{ik}\beta_{k} + f_0(\mathbf{s}_i)\,.
 \end{eqnarray*}
Then it tests two null hypotheses: $H_0:\bm{\beta}=(\beta_1,\ldots,\beta_K)^{\top}=\mathbf{0}$ and $H_0:f_0(\cdot)=0$ using the likelihood ratio test and the Wald test, respectively. If both null hypotheses are rejected, it proceeds to the second step. Second, it considers the full model with interactive effects, with the systematic structure (\ref{mod:celltypespecific_fixed}). It tests if any of the interactive effects $f_1(\cdot), \ldots, f_K(\cdot)$ are $0$ using the likelihood ratio test. Similar to spVC, CTSV \cite{yu2022identification} assumes a generalized additive model. However, CTSV has a different random structure and testing procedure. Specifically, CTSV assumes the random structure to be that $Y_i$ follows a ZINB distribution, and the systematic structure is written as
 \begin{eqnarray}
      g(\mu_i) = \beta_0 + \log\ell_i + \sum_{k=1}^{K} c_{ik}\beta_{k} + \sum_{k=1}^{K} c_{ik}f_{k}(\mathbf{s}_i),
 \end{eqnarray}
where $f_k(\mathbf{s}_i)$ is further assumed to be additive $f_k(\mathbf{s}_i) = f_{k1}(s_{i1}) + f_{k2}(s_{i2})$.
Unlike spVC's two-step testing procedure, CTSV directly tests if any of the interactive effects $f_1(\cdot), \ldots, f_K(\cdot)$ are $0$ using the Wald test. Same as spVC, C-SIDE \cite{cable2022cell} assumes the random structure to be Poisson. However, it has a different systematic structure from those of spVC and CTSV, allowing for random effects and additional non-linear transformations. C-SIDE uses a two-sided z-test.

For spatial-domain-marker SVG detection, the fixed-effect covariates of spot $i$ (i.e., $\mathbf{x}_i$) include the library size $\ell_i$ and the spatial domain, indicated by $\mathbf{d}_i = (d_{i1},\ldots,d_{iL}) \in \{0,1\}^{L}$, with $\sum_{l=1}^L d_{il} = 1$ (that is, $d_{il}=1$ if spot $i$ belongs to domain $l$). Then the systematic structure becomes 
 \begin{eqnarray}\label{mod:spaialde_fixed}
    g(\mu_i) = \beta_0 + \log \ell_i + \sum_{l=1}^{L} d_{il}\beta_{l}\,,
 \end{eqnarray}
where $\beta_{l}$ indicates the effect of spatial domain $l$ on $g(\mu_i)$, i.e., the transformed gene $i$'s true expression level.
For identifiability, constraints are needed for $\bm{\beta}=(\beta_1,\ldots,\beta_L)^{\top}$, e.g., $\beta_1 = 0$. The gene is a \textit{true spatial-domain-marker SVG} (i.e., a DEG across spatial domains) if $\bm{\beta} \neq \mathbf{0}$. 

SpaGCN and DESpace are two methods for spatial-domain-marker SVG detection. Following the systematic structure \eqref{mod:spaialde_fixed}, DESpace \cite{cai2024despace} assumes the gene expression $Y_i$ follows an NB distribution and tests the null hypothesis $H_0:\bm{\beta} = \mathbf{0}$ using the likelihood ratio test statistic, which follows an asymptotic chi-square distribution under the null hypothesis. For SpaGCN \cite{hu2021spagcn}, although it does not explicitly adopt a regression framework but instead uses the Wilcoxon rank-sum test on normalized gene expression levels (with the library size effect removed) to decide if a gene is  \rv{significantly more highly} expressed in one spatial domain compared to its neighboring spots, the Wilcoxon rank-sum is a score test statistic of the proportional odds ordinal logistic regression model. Hence, the test in SpaGCN can broadly be considered to follow the systematic structure (\ref{mod:spaialde_fixed}).




\subsubsection{Regression random-effect tests}

Unlike the nine methods that adopt dependence tests and the six methods that use regression fixed-effect tests, the last six of the 23 frequentist hypothesis-testing-based methods use regression random-effect tests: SpatialDE, nnSVG, SOMDE, SVCA, SPARK, and GPcounts.

All the six methods generalize the linear mixed model \eqref{mod:lm} for each gene, by encoding the spatial location $\mathbf{s}_i$ of spot $i$ as the random-effect covariate $\mathbf{z}_i$.
With $n$ spots, $\mathbf{z}_i = (z_{i1}, \ldots, z_{in})^{\top} \in \{0,1\}^n$ is a binary indicator vector with only one $1$ indicating spot $i$. Without loss of generality, we assume that $z_{ii} = 1$. Then, the corresponding random-effect vector $\bm{\gamma} = (\gamma_1, \ldots, \gamma_n)^{\top} \in \mathbb{R}^n$ has $\gamma_i$ indicating the random effect of $\mathbf{s}_i$. The covariance matrix of $\gamma_1, \ldots, \gamma_n$, $\Var(\bm{\gamma})$, is assumed to depend on the spatial proximity of $\mathbf{s}_1, \ldots, \mathbf{s}_n$. Accordingly, $\bm{\gamma}$ can be explicitly written as $\bm{\gamma}(\mathbf{s})$. When the null hypothesis $H_0: \Var(\bm{\gamma}(\mathbf{s})) = \mathbf{0}$ does not hold, the gene is a \textit{true overall SVG}.

As an early and exemplar method, for each gene SpatialDE \cite{svensson2018spatialde} assumes a linear random-effect model that only contains $\mathbf{z}_i$ as the random-effect covariates:
\begin{eqnarray}\label{mod:spatialde}
    Y_i = \beta_0 + \mathbf{z}_i^{\top}\bm{\gamma}(\mathbf{s}) + \epsilon_i,
\end{eqnarray} 
where the random errors $\epsilon_1,\ldots,\epsilon_n$ independently follow a Gaussian distribution $N(0, \delta)$, and the random effects $\bm{\gamma}(\mathbf{s})$ jointly follow a multivariate Gaussian distribution $\text{MVN}(\mathbf{0}, \sigma_s^2 \cdot \mathbf{K}(\mathbf{s}))$, with $\mathbf{K}(\mathbf{s}) = [K(\mathbf{s}_i, \mathbf{s}_j)]_{n\times n}$ specified by a kernel function $K(\cdot,\cdot)$ applied to the $n$ spatial locations. This model is essentially a Gaussian process. The null hypothesis then becomes $H_0:~ \sigma_s^2=0$.

Two methods, nnSVG \cite{weber2023nnsvg} and SOMDE \cite{hao2021somde}, also use the Gaussian process model as in SpatialDE but adopt more efficient computational algorithms. The other three methods extend the Gaussian process model. Specifically, SVCA \cite{arnol2019modeling} considers two additional variance terms to accommodate other covariates, such as cell-cell interactions. Instead of assuming that $Y_i$ follows a Gaussian distribution, SPARK \cite{sun2020statistical} and GPcounts \cite{bintayyashNonparametricModellingTemporal2021} assume Poisson and NB distributions, respectively, and generalize the linear mixed effect model (\ref{mod:spatialde}) to generalized linear mixed effect models.

All six methods use the likelihood ratio test with an asymptotic chi-square distribution under the null hypothesis.


\section{Discussion and future directions}\label{sec:challenge}
Below, we discuss the comparative advantages and trade-offs of existing SVG detection methods in terms of detection power, specificity, and scalability. Further, we outline two future directions for improving SVG detection methods: (1) \rv{accommodating differences among various SRT technologies and tissue types, as well as supporting multi-sample SRT data} and (2) enhancing statistical rigor and method validation. \rv{Last, we summarize the limitations of existing benchmark studies and provide guidance for future benchmark studies.} 

\subsection{Power, specificity, and scalability of SVG detection methods}

Among the 26 methods for detecting overall SVGs, the nine Euclidean-space-based, kernel-based methods define more specific spatial patterns for overall SVGs than the Euclidean-space-based, kernel-free methods and the graph-based methods. The statistical power of these kernel-based methods depends on the alignment between the pre-defined kernels' spatial patterns and the biologically relevant spatial patterns in SRT data. This alignment is often questionable because biologically relevant patterns are rarely as regular as the pre-defined kernel patterns. As a result, kernel-based methods may lose statistical power when this alignment is poor. To address this, some kernel-based methods, such as SPARK \cite{sun2020statistical}, incorporate multiple pre-defined kernels to capture a broader diversity of spatial patterns. Despite the potential loss of overall power, kernel-based methods are more specific and powerful for discovering overall SVGs whose spatial expression patterns align with the kernels compared to methods that do not use kernels. In short, there is a trade-off between overall power and specificity: methods targeting specific spatial patterns are less powerful at discovering other patterns, leading to a loss of overall power, but more powerful at discovering the targeted patterns with higher specificity. 

Four types of genes that exhibit interesting but non-global expression patterns can be easily missed by methods that detect overall SVGs. First, some genes of interest may only be \rv{highly} expressed in small regions of interest (ROIs) and can be overlooked by methods that do not distinguish small ROIs. For such genes, methods that detect spatial-domain-marker SVGs, such as SpaGCN \cite{hu2021spagcn}, might be more appropriate because these methods can identify small ROIs as spatial domains in the first step before identifying marker genes in the second step. Second, there are genes that exhibit spatial expression patterns within \rv{spatial domains}. These genes, referred to as \textit{spatial-domain-specific SVGs}, \rv{may be a subset of spatial-domain-marker SVGs because, although marker SVGs are highly expressed within a spatial domain, they may not exhibit informative spatial variation within that domain.} While no existing methods are specifically designed to detect these spatial-domain-specific SVGs, methods for detecting overall SVGs can likely be applied to specific spatial domains to discover these genes. \rv{Identifying spatial-domain-specific SVGs may help uncover spatial subdomains, capturing the internal variation within spatial domains. An alternative approach is to perform clustering within a spatial domain and then identify the resulting subdomain marker genes. However, these genes are not conceptually equivalent to those identified by an overall SVG detection method within the spatial domain, reflecting the same conceptual difference as that between overall SVGs and spatial-domain-marker SVGs.} Third, cell-type-specific SVGs might also be easily missed by methods that detect overall SVGs if the cell types of interest have small proportions. To address this, methods for detecting cell-type-specific SVGs have been developed, including CTSV \cite{yu2022identification}, C-SIDE \cite{cable2022cell}, and spVC \cite{yu2024spvc}. These methods rely on regression models, whose goodness-of-fit to SRT data remains to be explored. Ensuring a good fit is essential to avoid spurious discoveries. Fourth, some genes may exhibit sharp expression changes at tissue layer boundaries, which are too local to be detected by methods looking for overall SVGs. Belayer \cite{ma2022belayer} aims to detect such genes by examining gene expression change rates (gradients) in the 2D space. As a future direction, adding the accompanying H\&E image can help refine tissue boundaries, ensuring that the identified genes are interpretable. Moreover, if users have prior knowledge on some interesting genes that should be detected as SVGs, how to incorporate this knowledge into SVG detection to improve the specificity remains an open question.

The scalability of an SVG detection method is determined by the computational time of two steps: calculating a summary statistic for each gene and converting the summary statistic to a p-value based on a null distribution. The second step is only included in methods that provide statistical significance for the detected SVGs in a frequentist manner. In the first step of summary statistic calculation, computational time is measured in terms of \(n\), the number of spatial spots. For example, fitting a Gaussian process takes \(O(n^3)\) time in SpatialDE \cite{svensson2018spatialde} and SPARK \cite{sun2020statistical}, but this time is reduced to \(O(n)\) by using the nearest-neighbor Gaussian process approximation in nnSVG \cite{weber2023nnsvg}. By changing the modeling framework from a Gaussian process to a Pearson correlation between two similarity matrices, SPARK-X \cite{zhu2021spark} also achieves a fast computational time of \(O(n)\). In the second step of p-value calculation, computational time is fast if the summary statistic \rv{is a test statistic with a closed-form theoretical null distribution, as is the case for most methods that use regression tests}. However, \rv{methods that use test statistics require permutation to calculate the null distribution, making them computationally intensive}. To improve the scalability of a method, considerations can be put into expediting both steps. The first step can be accelerated using approximation algorithms. \rv{The number of permutations in the second step can be reduced through approaches such as p-value-free FDR control \cite{ge2021clipper}, parametric smoothing of the permutation-based null distribution \cite{song2021pseudotimede}, and adaptive strategies that use a large number of permutations only for potentially small p-values \cite{tiberi2023distinct}.}

\subsection{Future direction 1: Accommodating differences in SRT technology and \rv{tissue types, and supporting the analysis of multi-sample SRT data}}
There are two key differences among SRT technologies, yet most current SVG detection methods do not account for these differences in the pre-processing and modeling of SRT data. Below, we introduce these differences and discuss their potential impact on SVG detection.

1.~\textbf{Spatial resolution.} 
    Imaging-based SRT measures per transcript's spatial localization at a single-cell or subcellular resolution, whereas sequencing-based SRT captures data at a multicellular level with relatively coarse resolution. In imaging-based SRT data, detected SVGs might result from the irregular distribution of cell types, such as cancer cells spreading across tumor tissue or mixed cell types at the boundaries of tissue layers. In contrast, the coarse resolution of sequencing-based SRT likely results in smaller variance of gene expression levels across spatial spots compared to imaging-based SRT data. Consequently, the spatial resolution differences between the two types of SRT technologies necessitate different interpretations for the detected SVGs. However, most existing SVG detection methods do not distinguish between the SRT technologies but use the same approach to identify SVGs, indicating a need for improvement.

2.~\textbf{Positional randomness of spatial spots.}
Different SRT technologies use various strategies to record the positions of spatial spots, resulting in differences in positional randomness. For instance, technologies like Spatial Transcriptomics and 10x Visium capture transcripts in predefined rectilinear or hexagonal grids on micro-slides, leading to structured spatial spots that lack positional randomness. In contrast, technologies like Slide-seq, MERFISH, and SeqFISH capture transcripts wherever they are located, without using predefined grids, leading to unstructured spatial spots with positional randomness. This difference in spatial positional randomness necessitates different modeling strategies for spatial variance. For example, sepal \cite{andersson2021sepal} designs distinct spatial modeling frameworks?rectilinear, hexagonal, and unstructured spots?for Spatial Transcriptomics, 10x Visium, and Slide-seq, respectively. MERINGUE \cite{miller2021MERINGUE} and several other graph-based methods use Delaunay triangulation to account for the nonuniform density of unstructured spots. BOOST-HMI \cite{yang2024bayesian} uses the mark interaction model, an extension of the Ising model used in BOOST-MI, to address the irregular distribution of spatial spots for imaging-based SRT data. However, the majority of SVG detection methods have not considered the positional randomness differences among SRT technologies, potentially leading to biased detection of SVGs.

\rv{Moreover, the type of tissue plays a crucial role in SVG detection. Whether a tissue is morphologically structured or not should be considered when defining SVGs to ensure they are biologically meaningful. In well-structured tissues, such as the main olfactory bulb or cerebellum, meaningful SVGs are expected to exhibit regular expression patterns that align with the tissue's morphological structures. Therefore, kernel-based SVG detection methods, which identify SVGs based on regular kernel patterns, may be particularly relevant in these cases. In contrast, for tissues without clear morphological structures, such as tumors or diseased tissues, meaningful SVGs may display irregular patterns. In other words, the expression patterns of SVGs can differ significantly between well-structured and unstructured tissues. Consequently, SVG detection should account for the morphological characteristics of the tissue. Among existing methods, SpaGCN \cite{hu2021spagcn} follows this rationale by defining spatial domains through the joint analysis of SRT data and H\&E images, encouraging the identified spatial domains to reflect the underlying tissue structures. The SVGs identified by SpaGCN then serve as spatial-domain markers that explain these spatial domains. However, it remains an open question whether incorporating H\&E images into overall SVG detection is necessary, especially if overall SVGs are primarily used as pre-screened features for further analysis and computational efficiency is a primary concern. For cell-type-specific SVG detection, H\&E images have not yet been incorporated into existing methods. This raises an interesting future research question: how do cell types interact with morphological structures?}

\rv{Another open question and future direction is the detection of SVGs from multi-sample SRT data, such as consecutive slices of a single tissue or tissue slices from different patients. Existing methods typically detect SVGs from a single tissue slice; however, intuitively, SVGs should be meaningful features that can be compared across multiple tissue slices. The challenge, therefore, is how to effectively identify SVGs from multi-sample SRT data to enable downstream multi-sample comparisons (e.g., normal vs. diseased tissue). In multi-sample SRT data of well-structured tissues, a common strategy is to align the tissue slices into a common 2D coordinate system \cite{clifton2023stalign, jones2023alignment, zhang2023molecularly, preibisch2009globally}. This alignment facilitates the detection of SVGs. However, for unstructured tissues (e.g., tumors), alignment may not be feasible or reasonable, necessitating new strategies for SVG detection that can summarize the common, meaningful spatial variations in multi-sample SRT data of unstructured tissues.}

\subsection{Future direction 2: Enhancing statistical rigor and method validation}

Similar to single-cell RNA-seq data analysis \cite{zhang2019valid, neufeld2024inference, song2023clusterde}, SRT data analysis faces the "double-dipping" challenge: the same data are analyzed more than once, and the final statistical tests rely on variables inferred from the same data in previous steps, leading to a "confirmation bias". This issue is prominent in spatial-domain-marker SVG detection, where the genes involved in identifying spatial domains would inherently be identified as spatial-domain markers in a subsequent step, even if they do not exhibit significant expression changes between spatial domains. Therefore, strategies are needed to remove false-positive discoveries resulting from double-dipping, such as ClusterDE \cite{song2023clusterde}.

Moreover, as many SVG detection methods use complex algorithms with implicit assumptions, interpretable ways for sanity checks on the detected SVGs are essential. For example, using \textit{in silico} negative control SRT data can help users test SVG detection methods and identify spurious discoveries. Additionally, fast visualization tools will assist users in interpreting the top detected SVGs, providing a clearer understanding of the results and enhancing the reliability of the findings.

\begin{figure}[htbp]
    \centering    \includegraphics[width=0.6\textwidth]{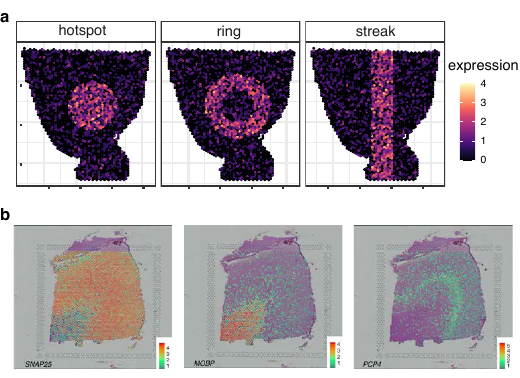}
    \caption{\textbf{Synthetic spatial patterns oversimplify the spatial patterns observed in real SRT data.} \textbf{a}, Representative spatial patterns used in synthetic SRT data for evaluating SVG detection methods such as Trendsceek \cite{edsgard2018trendsceek}, SpatialDE \cite{svensson2018spatialde}, and others.
    \textbf{b}, Spatial patterns shown in the 10x Visium dataset \cite{maynard2021transcriptome} profiling a dorsal lateral prefrontal cortex sample. The color indicates the $\log$ transformed expression for genes \textit{SNAP25}, \textit{MOBP}, and \textit{PCP4}. }
    \label{fig:pattern}
\end{figure}

No method can be optimal in every aspect, making method choice for users' goals crucial. Users should select methods that capture the spatial patterns of interest, ensuring the method's strengths align with their research needs. Consequently, the diversity of methods is indispensable to cater to various user needs, and method validation and benchmarking are necessary for users to choose the appropriate methods. However, benchmarking is challenging due to the lack of well-annotated datasets with SVG ground truths. As a result, methods often justify their effectiveness indirectly, which may or may not reflect the biological questions users seek answers for. Early methods like SpatialDE and Trendsceek validate their detected SVGs by visual inspection. Another prevalent strategy is to use synthetic datasets with artificial spatial patterns, such as hotspots, streaks, and rings (Fig.~\ref{fig:pattern}a), to evaluate methods' detection power. However, these artificial patterns are oversimplified and might not be biologically relevant (Fig.~\ref{fig:pattern}b). Hence, benchmarking needs well-annotated SRT datasets from diverse tissues (e.g., structured tissues like the brain and unstructured tissues like tumors) and realistic SRT data simulators \cite{zhu2023srtsim, song2024scdesign3}, not only for comprehensive method validation but also for the potential development of supervised methods. We curated the SRT datasets used by the existing SVG detection methods in Supplementary File. Future research is needed to enhance the realism, comprehensiveness, and scalability of SRT data simulators.

\subsection{Guidance for future benchmark studies}\label{sec:benchmark}

\rv{To date, three benchmark studies have been conducted to evaluate the performance of selected SVG detection methods \cite{li2023benchmarking, chen2024evaluating, charitakis2023disparities}. These studies highlight the complexity of the SVG detection task and the variability in the performance of existing methods. In Supplementary Table~1, we summarize the three benchmark studies, detailing their choices of methods, benchmark datasets, evaluation metrics, and conclusions.}

\rv{A limitation of the existing benchmark studies is their lack of consideration for the conceptual differences among SVG detection methods (in our categorization: overall SVGs, cell-type-specific SVGs, and spatial-domain-marker SVGs). This limitation can lead to unnecessary comparisons that are essentially like comparing apples to oranges. For instance, two benchmark studies \cite{li2023benchmarking, charitakis2023disparities} compared overall SVG detection methods (such as SpatialDE and SPARK) with spatial-domain-marker SVG detection methods (such as SpaGCN), using the number of SVGs detected as the evaluation metric. However, these two types of methods have distinct goals: overall SVG detection methods are intended to serve as a preprocessing step for feature selection and are inherently more likely to detect a greater number of genes as SVGs, while spatial-domain-marker SVG detection methods are specific to pre-defined spatial domains and tend to identify fewer genes as spatial domain markers. Conceptually, the number of detected genes is not a fair metric for comparing these two types of methods. Indeed, this anticipated result was confirmed in both benchmark studies. Had this conceptual categorization been recognized before the benchmark studies were conducted, computational resources could have been conserved by avoiding such comparisons.}

\rv{A second limitation of the existing benchmark studies is that, although they compared SVG detection methods using diverse methodologies, their conclusions did not adequately summarize or emphasize how different core methodologies (e.g., graph-based or kernel-based approaches) affect the methods' performance within the benchmarks. This limitation means that the impact of fundamental methodological choices on the effectiveness of SVG detection remains underexplored, potentially limiting the insights that could guide future method development.}

\rv{A third limitation of the existing benchmark studies is their varied selection of SRT datasets and benchmarking scenarios, leading to evaluation results and conclusions that are not fully comparable. Using limited benchmarking scenarios may favor methods that perform well under specific conditions while overlooking their robustness across a broader range of spatial expression patterns. Consequently, without a consensus on benchmark design, the conclusions drawn from these studies may not provide a comprehensive assessment of the strengths and weaknesses of the methods being compared.}


\rv{In light of the three identified limitations in existing benchmark studies, we offer three recommendations for future benchmark studies, guided by our conceptual categorization of SVG detection methods. Recognizing that the definitions of SVGs differ among the three conceptual categories (overall SVGs, cell-type-specific SVGs, and spatial-domain-marker SVGs), it is both unfair and uninformative to compare methods across these categories. Therefore, our first recommendation is that future benchmark studies should evaluate methods within each conceptual category, providing more accurate and insightful assessments of SVG detection methods, and better guiding researchers in selecting the most appropriate methods for their specific needs.}

\rv{Our second recommendation is to benchmark distinct methodologies aimed at achieving the same detection goal. This approach is valuable for method developers, as it provides a clearer understanding of the sensitivity and specificity of different methodologies, helping to avoid reinventing the wheel or encountering unknown pitfalls in future method development. For example, in the context of overall SVG detection, it would be insightful to compare the performance of dependence tests (with a general independence null hypothesis) against regression tests (with a model-based null hypothesis). Another question worth investigating is whether graph-based methods using Delaunay triangulation, such as MERINGUE and Binspect, can better capture cellular adjacency compared to KNN-based methods like SpaGene and Hotspot in tissues where cells are non-uniformly arranged.}

\rv{Our third recommendation is to include SRT datasets from diverse tissue types and technologies, along with comprehensive and carefully designed simulations, to ensure broad coverage of informative spatial expression patterns. While this practice is challenging, it is essential to avoid benchmark conclusions that may biasedly favor certain types of SVG detection methods. For instance, if SRT data is simulated based on a specific kernel pattern, benchmark results will inherently favor SVG methods tailored to that kernel. Similarly, if the simulation design includes only a limited range of spatial expression patterns, methods using dependence tests---which operate under a general independence null hypothesis and can detect SVGs with diverse patterns---may be unfairly disfavored. We curated the SRT datasets used by the existing SVG detection methods in Supplementary File. Looking forward, it is essential for the community to collaboratively curate a benchmark database to ensure fair benchmarking and foster future method development.}

\section{Conclusion}{\label{sec:conclusion}}



In this article, we review 33 peer-reviewed SVG detection methods for SRT data, focusing on the SVG definitions, detection methodologies, and biological implications. To avoid the ambiguity of terminology, we classify the 33 methods into three main categories based on the SVGs they detect: overall SVGs, cell-type-specific SVGs, and spatial-domain-marker SVGs (Fig.~\ref{fig:catecartoon}--\ref{fig:decisiontree}; Table~\ref{tab:cate}). We summarize these methods in Tables~\ref{tab:tab_model}--\ref{tab:tab_test}, including input data, modeling framework, and availability of statistical significance. Next, we introduce the downstream applications of SVGs, including identifying spatial domains with marker genes and spatial patterns of cell types or states. Furthermore, from a statistical perspective, we divide the frequentist hypothesis tests used in 23 methods into three types: dependence tests, regression fixed-effect tests, and regression random-effect tests (Table~\ref{tab:cate}). We summarize how the three types of hypothesis tests are used to detect SVGs and discuss how they differ from each other regarding the generality and specificity for SVG detection.
Finally, we discuss the comparative advantages of existing SVG detection methods and offer insights into the future directions for improvement. 

\section*{Competing interests}
The authors declare no competing interests. \vspace*{-12pt}

\section*{Acknowledgements}
The authors appreciate the comments and feedback from the members of the Junction of Statistics and Biology at UCLA (\url{http://jsb.ucla.edu}). \vspace*{-12pt}

\section*{Funding}
This work was supported by the following grants: National Science Foundation DBI-1846216 and DMS-2113754, NIH/NIGMS R01GM120507 and R35GM140888, Johnson $\&$ Johnson WiSTEM2D Award, Sloan Research Fellowship, UCLA David Geffen School of Medicine W.M. Keck Foundation Junior Faculty Award, and Chan-Zuckerberg Initiative Single-Cell Biology Data Insights [Silicon Valley Community Foundation Grant Number: 2022-249355] (to J.J.L.). J.J.L. was a fellow at the Radcliffe Institute for Advanced Study at Harvard University in 2022?2023 while she was writing this paper. \vspace*{-12pt}

\newpage
\clearpage
\section*{Figures}

\clearpage

\clearpage
\section*{Tables}\label{sec:tab}

\begin{table}[htbp]
\caption{\textbf{Classification of 33 SVG detection methods along with statistical inference types}}\label{tab:cate}
\begin{tabularx}{\textwidth}{|X|p{1.25in}|p{1.2in}|X|X|}
\cmidrule{1-5}
\multicolumn{2}{|l|}{\diagbox[width=6.7cm]{{Inference types}}{{Categories}}} & {Overall SVGs} & \makecell[l]{Cell-type- \\ specific SVGs} & \makecell[l]{Spatial-domain- \\ marker SVGs} \\
\cmidrule{1-5}
\multirow{18}{*}{\makecell[l]{Frequentist \\ inference}} & \multirow{9}{*}{\makecell[l]{Dependence \\ tests}} & SPARK-X &  & \\
& & SMASH & & \\
& & singlecellHaystack & & \\
& & Trendsceek & & \\
& & HEARTSVG & & \\
& & Hotspot & & \\
& & MERINGUE & &  \\
& & SpaGene & & \\
& & BinSpect & & \\
& & scGCO & & \\
& & RayleighSelection & & \\
\cline{2-5}
& \multirow{3}{*}{\makecell[l]{Regression \\ fixed-effect tests}} & & CTSV & SpaGCN \\
& & SPADE & C-SIDE & DESpace \\
& & & spVC & \\
\cline{2-5}
& \multirow{6}{*}{\makecell[l]{Regression \\ random-effect tests}} & SpatialDE & & \\
& & nnSVG & & \\
& & SOMDE & & \\
& & SVCA &  &  \\
& & SPARK & & \\
& & GPcounts & & \\
\cmidrule{1-5}
\multicolumn{2}{|l|}{} & BOOST-GP & & \\
\multicolumn{2}{|l|}{Bayesian inference} & BOOST-MI &  & \\
\multicolumn{2}{|l|}{} & BOOST-HMI & & \\
\cmidrule{1-5}
\multicolumn{2}{|l|}{} & MULTILAYER & & \\
\multicolumn{2}{|l|}{} & sepal & & \\
\multicolumn{2}{|l|}{No statistical inference} & BSP$^1$ &  & \\
\multicolumn{2}{|l|}{} & PROST$^2$ & & \\
\multicolumn{2}{|l|}{} & HRG & & \\
\multicolumn{2}{|l|}{} & SVGbit & & \\
\multicolumn{2}{|l|}{} & SINFONIA & & \\
\cmidrule{1-5}
\end{tabularx}
\footnotesize{$^1$ Although BSP attempts to perform frequentist inference by defining a null distribution for its test statistic, the null distribution is improperly defined as a distribution fitted to the test statistic values of all genes, implying that all genes are non-SVGs. Hence, we do not label BSP as a method with statistical inference in the table.
}\\
\footnotesize{$^2$ PROST performs frequentist statistical tests for Moran?s I, instead of the PROST index it uses to rank genes as SVGs.}\\
\end{table}

\begin{table}[htbp]
    \centering       
    \captionof{table}{\textbf{Methodological traits of 33 SVG detection methods.}}
\includegraphics[width=\linewidth]{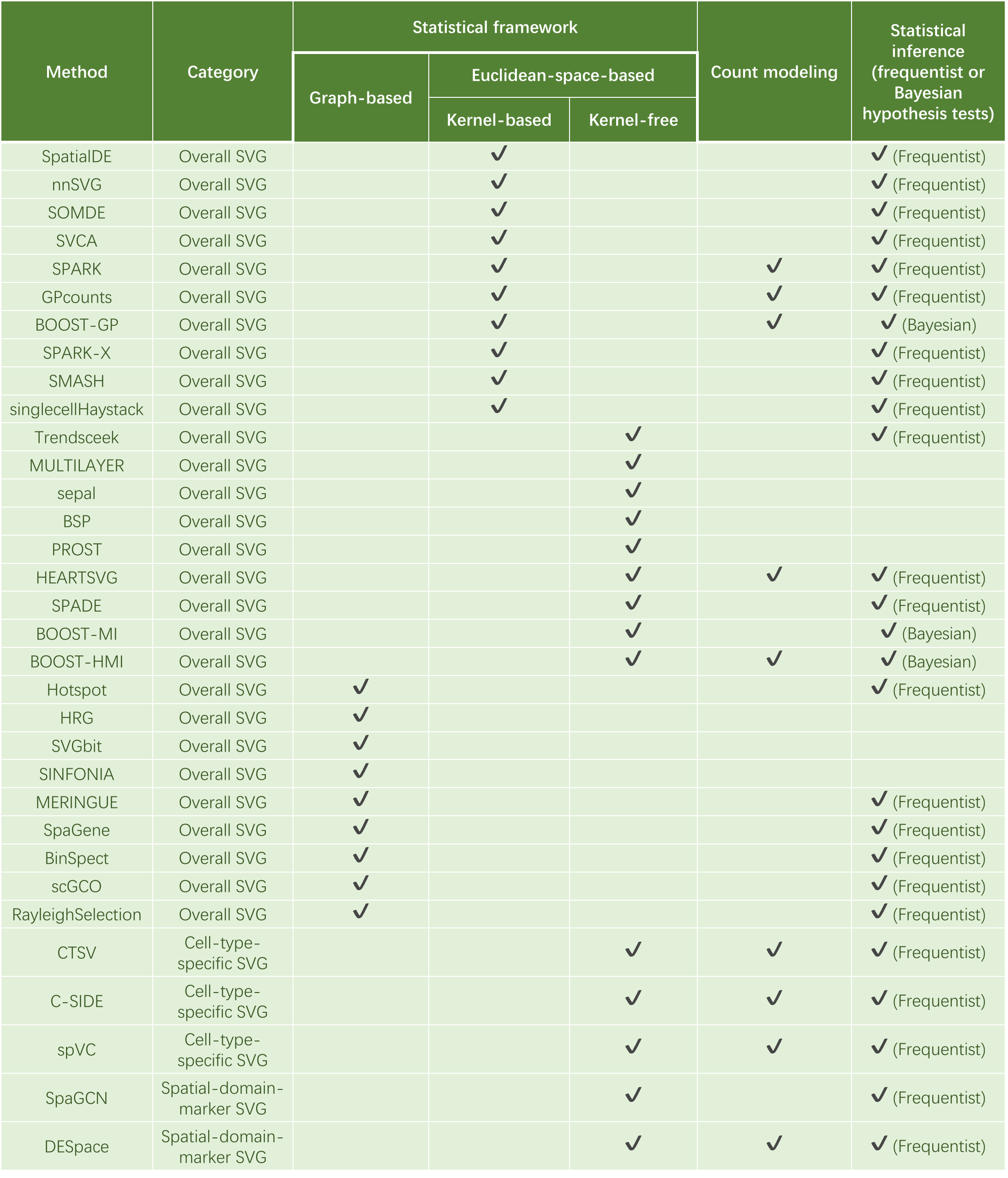}
  \label{tab:tab_model}
\end{table}

\begin{table}[htbp]
    \centering     
      \captionof{table}{\textbf{Characteristics of 23 frequentist-hypothesis-tests-based SVG detection methods.}$^{1,2}$}
    \label{tab:tab_test}
\includegraphics[width=\textwidth]{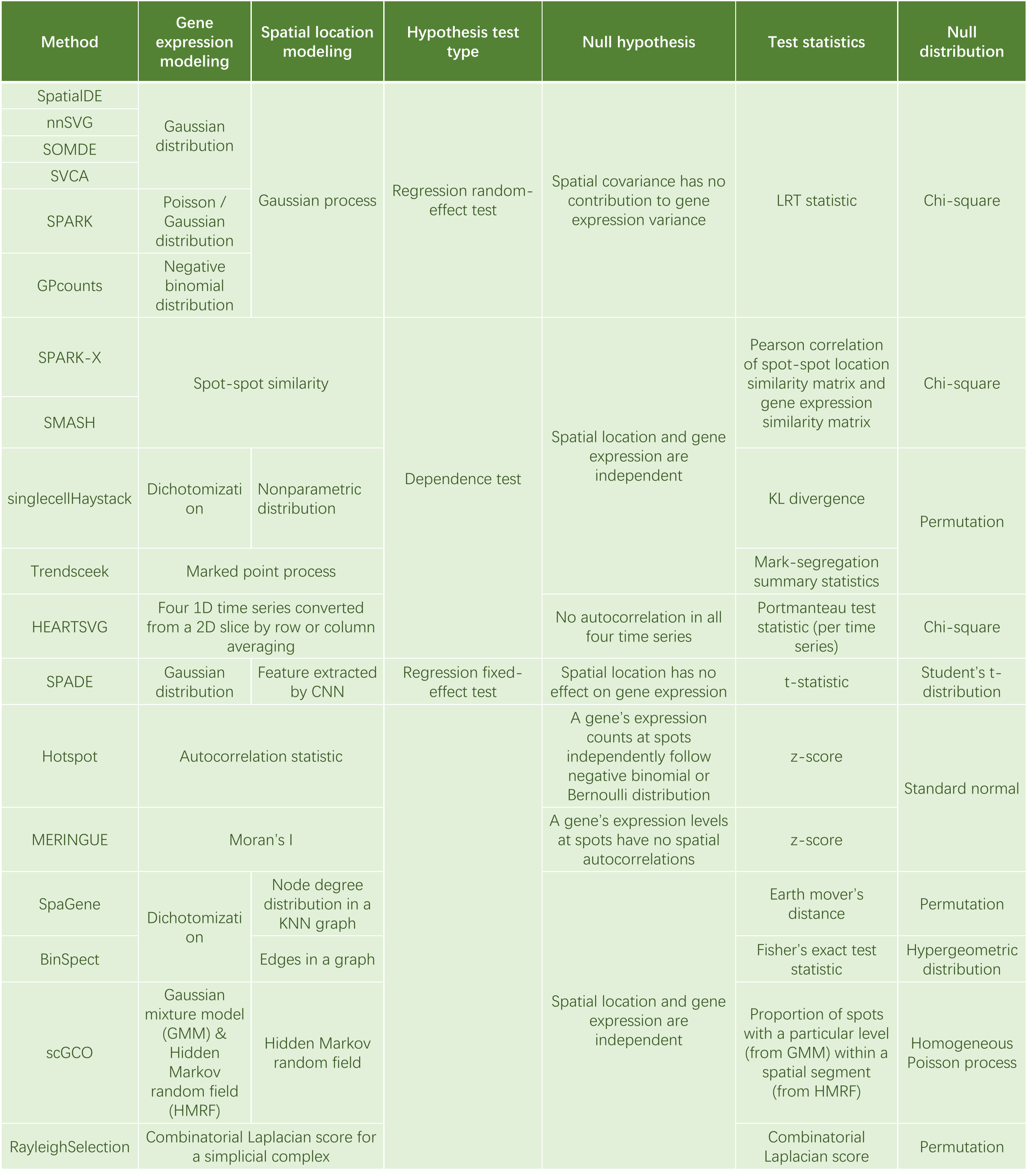}
\begin{flushleft}
\footnotesize{$^1$ The 23 SVG detection methods using frequentist statistical hypothesis tests (Table~\ref{tab:tab_model}) are included in this table.}\\
\footnotesize{$^2$ The table is continued on the next page.}
\end{flushleft}
\end{table}

\begin{table}[htbp]
    \centering     
      \captionof*{table}{\textbf{Table 3: Characteristics of 23 frequentist-hypothesis-tests-based SVG detection methods. (continued)}}
\includegraphics[width=\textwidth]{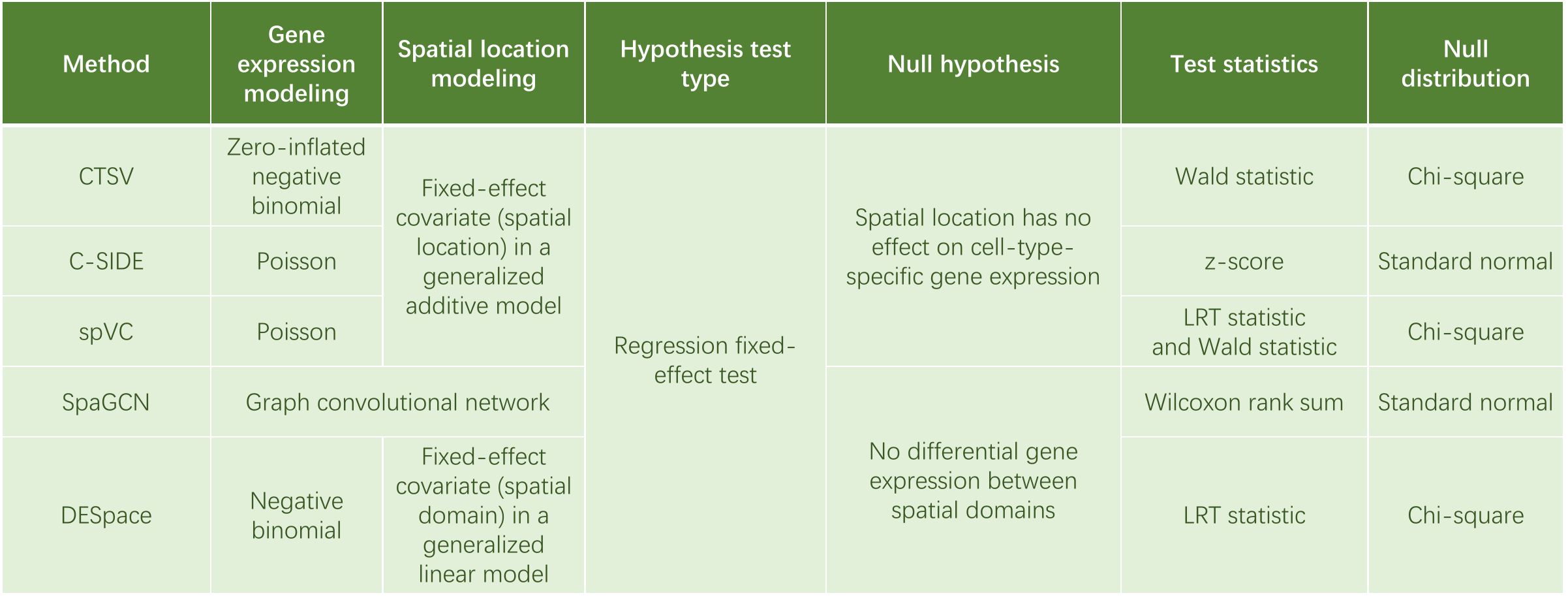}
\end{table}

\clearpage
\begin{table}[htbp]
\centering
\caption{Definitions of concepts for SVGs and frequentist statistical tests}
\begin{tabular}{|p{2cm}|l|p{8cm}|}
\hline
 & \textbf{Concept} & \textbf{Definition} \\
\hline
\multirow{3}{*}{\textbf{SVGs}} & Overall SVGs & Genes that exhibit non-random spatial expression patterns. This is the most general category of SVGs detected using only SRT data without incorporating external information such as spatial domains or cell types. \\
\cline{2-3}
& Cell-type-specific SVGs & Genes that exhibit non-random spatial expression patterns within a cell type. These genes are detected using both SRT data and external cell-type annotations for the spatial spots. \\
\cline{2-3}
& Spatial-domain-marker SVGs & Genes that exhibit significantly higher expression in a spatial domain compared to other domains. These genes are detected using SRT data and spatial domains, which are usually detected from the same SRT data. \\
\hline
\multirow{3}{*}{\makecell[l]{\textbf{Frequentist} \\\textbf{Tests}}} & Dependent Test & Examines the dependence between a gene?s expression level and the spatial location. \\
\cline{2-3}
& Fixed-effect Tests & Examines whether some or all of the fixed-effect covariates $\mathbf{x}_i$ contribute to the mean of the response variable, i.e., a gene's expression. \\
\cline{2-3}
& Random-effect Tests & Examines whether the random-effect covariates $\mathbf{z}_i$ contribute to the variance of the response variable, i.e., a gene's expression. \\
\hline
\end{tabular}
\label{tab:concepts_summary}
\end{table}

\clearpage

\begin{table}[htbp]
\centering
\caption{Mathematical notations and interpretations}
\begin{tabular}{|l|p{8cm}|}
\hline
\textbf{Symbol} & \textbf{Interpretation} \\
\hline
$n \in \mathbb{Z}^{+}$ & Number of spatial spots \\
\hline
$y_{i} \in \mathbb{R}$ & Expression level of a gene at spot $i$ \\
\hline
$Y_{i} \in \mathbb{R}$ & Random variable representing the expression level of a gene at spot $i$ \\
\hline
$\mathbf{Y} = (Y_1, \ldots, Y_n)^\top \in \mathbb{R}^{n}$ & Random vector of a gene's expression levels at all $n$ spatial spots \\
\hline
$\mathbf{s}_i = (s_{i1}, s_{i2})^\top \in \mathbb{R}^{2}$ & 2D spatial location of spot $i$ \\
\hline
$\mathbf{s} = \left[\mathbf{s}_{1}, \ldots, \mathbf{s}_{n}\right]^\top \in \mathbb{R}^{n \times 2}$ & Matrix of spatial coordinates for all $n$ spots \\
\hline
$\mathbf{d}_i = (d_{i1}, \ldots, d_{iL})^\top \in \{0, 1\}^{L}$ & Spatial-domain indicator vector for spot $i$ \\
\hline
$\mathbf{c}_i = (c_{i1}, \ldots, c_{iK})^\top \in [0, 1]^{K}$ & Cell-type proportion vector for spot $i$ \\
\hline
$\mathbf{x}_i \in \mathbb{R}^{p}$ & Fixed-effect covariates of spot $i$ \\
\hline
$\mathbf{z}_i \in \mathbb{R}^{q}$ & Random-effect covariates of spot $i$ \\
\hline
$\epsilon_i \in \mathbb{R} $ & Random measurement error at spot $i$ \\
\hline
$\bm{\epsilon} = (\epsilon_1, \ldots, \epsilon_n)^\top \in \mathbb{R}^n$ & Vector of random measurement errors with \newline assumed independence \\
\hline
$\beta_0\in \mathbb{R}$ & (Fixed) intercept in the linear mixed model \\
\hline
$\bm{\beta} \in \mathbb{R}^{p}$ & Fixed effects in the linear mixed model \\
\hline
$\bm{\gamma} \in \mathbb{R}^{q}$ & Random effects in the linear mixed model \\
\hline
$\bm{\Sigma} \in \mathbb{R}^{q \times q}$ & Covariance matrix of random effects $\bm{\gamma}$ \\
\hline
\end{tabular}
\label{tab:notations}
\end{table}

\newpage

\end{document}